\documentclass[preprint,12pt]{aastex}

\def\msun{{\rm\,M_\odot}} 
\def\lsun{{\rm\,L_\odot}}
\def\zsun{{\rm\,Z_\odot}}
\newcommand{\etal}{et al.\ }

\newcommand{\lya}{Ly$\alpha$ }

\def\h2{${\rm\,H_2}$}
\def\muG{\rm \mu{G}}

\slugcomment{Submitted to ApJ}

\begin{document}

\title{The Universe Was Reionized Twice}

\author{Renyue Cen\altaffilmark{1}}

\altaffiltext{1} {Princeton University Observatory, 
Princeton University, Princeton, NJ 08544; cen@astro.princeton.edu}

\accepted{ }

\begin{abstract}

We show the universe was reionized twice,
first at $z\sim 15-16$ and second at $z\sim 6$.
Such an outcome appears inevitable,
when normalizing to two well determined observational measurements,
namely, the epoch of the final cosmological reionization at $z\sim 6$
and the density fluctuations at $z\sim 6$, which in turn 
are tightly constrained by \lya forest observations at $z\sim 3$.
These two observations most importantly fix
the product of star formation efficiency 
and ionizing photon escape fraction from galaxies at high redshift.
The only major assumption made is that the initial mass function
of metal-free, Population III stars is top-heavy.

To the extent that the relative star formation efficiencies
in gaseous minihalos with \h2 cooling
and large halos with atomic cooling
at high redshift are unknown,
the primary source for the first reionization is still uncertain.
If star formation efficiency in minihalos 
is at least 10\% of that in large halos,
then Pop III stars in the minihalos may be largely responsible
for the first reionization;
otherwise, the first reionization 
will be attributable largely to
Pop III stars in large halos.
In the former case, \h2 cooling in minihalos
is necessarily efficient.
We show that 
gas in minihalos can be cooled efficiently by \h2 molecules
and star formation can continue to take place
largely unimpeded throughout the first reionization period,
as long as gas is able to accumulate in them.
This comes about thanks to  
two new mechanisms for generating
a high X-ray background during the Pop III era,
put forth here,
namely, X-ray emission from the cooling energy of 
Pop III supernova blast waves and 
from miniquasars powered by Pop III black holes.
Consequently, \h2 formation in the cores of minihalos
is significantly induced to be able to counteract
the destruction by Lyman-Werner photons
produced by the same Pop III stars.
In addition, an important process for 
producing a large number of \h2 molecules in relic HII
regions of Pop III galaxies, first 
pointed out by Ricotti, Gnedin, \& Shull,
is quantified here.
It is shown that \h2 molecules produced 
by this process may overwhelm the Lyman-Werner photons 
produced by stars in the same Pop III galaxies.
As a result, the Lyman-Werner background may never 
build up in the first place during the Pop III era.

The long cosmological reionization and reheating history is complex.
From $z\sim 30$ Pop III stars gradually heat up and ionize
the intergalactic medium, 
completing the first reionization at $z\sim 15-16$,
followed by a brief period of $\Delta z\sim 1$,
during which the intergalactic medium stays completely 
ionized due to sustained ionizing
photon emission from concomitant Pop III galaxies.
The transition from Pop III stars to Pop II stars
at $z\sim 13$ suddenly reduces, by a factor of $\sim 10$, 
ionizing photon emission rate,
causing hydrogen to rapidly recombine,
marking the second cosmological recombination.
From $z\sim 13$ to $z=6$,
Compton cooling by the cosmic microwave background
and photoheating by the stars self-regulate 
the Jeans mass and the star formation rate,
giving rise to a mean temperature of the intergalactic medium
maintained nearly at constant of $\sim 10^4~$K.
Meanwhile, recombination and photoionization 
balance one another such that
the intergalactic medium stays largely ionized
during this stage with $n_{\rm HII}/n_{\rm H}\ge 0.6$.
Most of the star formation in this period occurs in
large halos with dominant atomic line cooling.

We discuss a wide range of implications and possible
tests for this new reionization picture.
In particular, the Thomson scattering optical depth is increased to
$0.10\pm 0.03$, compared to $0.027$ 
for the case of only one rapid reionization at $z=6$.
Upcoming Microwave Anisotropy Probe observation of the
polarization of the cosmic microwave background
should be able to distinguish between these two scenarios.

\end{abstract}


\section{Introduction}

The conventional view is that the universe becomes reionized
at some point in the redshift range $z=6-10$,
when the UV emission rate in 
galaxies with virial temperatures greater than 
$\sim 10^4$K (where hydrogen atoms are efficient coolants)
exceeds the overall recombination rate (e.g., Barkana \& Loeb 2001; Madau 2002).
At the time of this writing
the redshift range has been narrowed to a point
at $z\sim 6$, as suggested by recent
observations of high redshift 
quasars from the Sloan Digital Sky Survey (SDSS)
(e.g., Fan \etal 2001; Becker \etal 2001, Barkana 2001; 
Cen \& McDonald 2002; Litz \etal 2002).

In this paper we present a new scenario.
In the context of the standard cold dark matter cosmological model
we show that the universe was first reionized at $z=15-16$ by
Population III (Pop III) stars and second reionized at $z=6$.
Following the first reionization, 
the transition from Pop III
stars to Pop II stars occurs.
At this point photoionization becomes insufficient to counterbalance the rapid
recombination process and 
the intergalactic medium (IGM)
recombines to become opaque to \lya and ionizing photons, again!
As time progresses,
with increased density fluctuations and the nonlinear mass scale
star formation rate gradually picks up.
At $z\sim 6$ the global star formation rate 
exceeds the global recombination rate
and the universe is completely reionized for the second time, and for all.
From the first cosmological reionization through
the second cosmological reionization 
the mean temperature of the IGM is maintained at $\sim 10^4~$K 
balanced between Compton cooling and photo-heating 
and hydrogen is more than half ionized balanced
between recombination and photoionization.

We show that this new reionization picture 
is inevitable, {\it as long as Pop III initial mass function (IMF) 
is top-heavy},
thanks to a new, powerful constraint 
placed on the product of star formation efficiency
and ionizing photon escape fraction from galaxies
at high redshift, which otherwise would be unknown.
This constraint comes from two solid pieces of observations:
1) the universe is required to be reionized at $z\sim 6$
and 2) the density fluctuation in the universe
at $z\sim 6$ is well determined by 
the same small-scale power traced by the observed
\lya forest observed at $z\sim 3$.

While it is clear that the ionizing sources for the second
cosmological reionization are stars in large galaxies with
efficient atomic line cooling,
the ionizing sources for the first 
cosmological reionization are uncertain 
to the extent that we do not know 
what the relative star formation efficiencies
in large halos and minihalos are.
We define ``minihalos" as those whose
virial temperature is less than $\sim 8\times 10^3$K where
only \h2 cooling is possible in the absence of metals,
and large halos as those with 
virial temperature above $\sim 8\times 10^3$K
capable of cooling via atomic lines.
We show that,
if star formation efficiency in minihalos with \h2 cooling
is not more than
a factor of ten less efficient than in large halos with atomic cooling,
then Pop III stars in minihalos may be largely responsible
for the first reionization.
Conversely, the first reionization may be attributable largely to
Pop III stars in large halos.

In order to enable efficient \h2 cooling in minihalos
it is necessary to maintain an adequate level of 
\h2 molecule fraction within their cores.
As is well known, \h2 molecules are  
fragile and easily destroyed by photons 
in the Lyman-Werner (LW) bands
($11.18-13.6$eV; Field \etal 1966; Stecher \& Williams 1967),
to which the universe
is largely transparent.
Primeval \h2 molecules have been completely
destroyed well before enough ionizing photons are produced
to reionize the universe 
(Gnedin \& Ostriker 1997; 
Haiman, Rees, \& Loeb 1997;
Tegmark \etal 1997).
Therefore, an adequate production rate of \h2 molecules
is required to counteract the destruction rate to keep
\h2 fraction at a useful level.
A sufficiently high X-ray background at high redshift
could serve as a requisite catalyst for 
forming \h2 molecules by deeply penetrating into and 
generating a sufficient number of free electrons 
in the cores of minihalos (Haiman, Rees, \& Loeb 1996;
Haiman, Abel, \& Rees 2000;  Ricotti, Gnedin, \& Shull 2001, 
RGS hereafter; Glover \& Bland 2002).

We put forth two new mechanisms for generating
a high X-ray background during the Pop III era.
We point out that at $z=13-20$
Pop III supernova remnants 
and miniquasars powered by Pop III black holes
are efficient X-ray emitters.
The much higher density of the interstellar
medium at high redshift
results in a much more rapid and earlier cooling phase of the supernova
blast waves, emitting a large fraction of
the cooling energy in X-ray energy ($\sim 1$keV). 
[In contrast,
supernova blast waves in the local galaxies 
are known not to cool efficiently until their temperatures
have dropped well below the X-ray regime (e.g., Cox 1972; 
Chevalier 1974,1982)].
Miniquasars powered by 
Pop III black holes 
of mass 
$\sim 10-100\msun$ 
are expected to
emit a large fraction of
the radiation in X-rays at $\ge 1~$keV.
The combined X-ray emission from these two sources
is sufficient to enable efficient \h2 
formation and cooling in the cores of minihalo galaxies 
in the redshift period in question.

In addition, an important process for 
producing a large number of \h2 molecules in relic HII
regions of Pop III galaxies is quantified.
It is shown that \h2 molecules produced 
by this process may overwhelm the Lyman-Werner photons 
produced by stars in the same Pop III galaxies.
As a result, the Lyman-Werner background may have never 
built up in the first place during the Pop III era.
In combination, we suggest that 
star formation in gas-rich minihalos can continue to take place
largely unimpeded throughout the first reionization period.

The outline of this paper is as follows.
Two new sources of X-ray radiation,
namely, the Pop III supernova cooling radiation
and X-ray emission from miniquasars are quantified in \S 2.
In \S 3 we quantify the \h2 formation in relic HII regions
of Pop III galaxies.
In \S 4 we compute in detail
the evolution of the IGM from $z\gg 20$ to $z=6$.
We discuss possible implications and significant
consequences and tests of this new scenario in \S 5,
and conclude in \S 6.
Throughout a spatially flat cold dark matter cosmological model 
with $\Omega_M=0.25$, $\Omega_b=0.04$, $\Lambda=0.75$,
$H_0=72$km/s/Mpc and $\sigma_8=0.8$ is adopted
(e.g., Bahcall \etal 2002).

\section{X-ray Emission From Pop III Supernovae and Miniquasars, and Molecular
Hydrogen Formation in Minihalos}

The important role that \h2 molecules play in 
the collapse of gas clouds to form Pop III stars
and the initial mass function of Pop III stars
have a long history of investigation
(Saslaw \& Zipoy 1967;
Hirasawa, Aizu, \& Taketani 1969;
Takeda, Sato, \& Matsuda 1969;
Hutchins 1976;
Silk 1977;
Hartquist \& Cameron 1977;
Shchekinov \& Edelman 1978;
Yoshii \& Sabano 1979;
Tohline 1980; 
Carlberg 1981; 
Lepp \& Shull 1984;
Yoshii \& Saio 1986;
Lahav 1986;
Stahler 1986;
Shapiro \& Kang 1987;
Uehara \etal 1996;
Haiman, Thoul, \& Loeb 1996;
Padoan, Raul, \& Jones 1997;
Nakamura \& Umemura 1999,2001,2002;
Larson 1995,2000;
Abel \etal 1998;
Abel, Bryan, \& Norman 1999,2000,2002;
Bromm, Coppi, \& Larson 1999, 2002;
Fuller \& Couchman 2000;
Machacek, Bryan, \& Abel 2001).
Many authors have examined 
the possibility of Pop III stars in minihalos
reionizing the universe (e.g., Couchman \& Rees 1986;
Fukugita \& Kawasaki 1994; Haiman \& Loeb 1997)
or partially reionizing
the universe (e.g., Gnedin \& Ostriker 1997;
Gnedin 2000a; RGS).
The primary difficulty for Pop III stars in minihalos 
to reionize
the universe has by now been realized to 
be the destruction 
of \h2 molecules by photons in the LW bands
produced by the same Pop III stars,
long before the completion 
of reionization process
(Gnedin \& Ostriker 1997; 
Haiman, Rees, \& Loeb 1997;
Tegmark \etal 1997);
\h2 photodissociation time becomes shorter
than the Hubble time when the ionizing
radiation intensity at the Lyman limit 
reaches $\sim 10^{-24}$ erg/cm$^2$/sec/hz/sr at the
redshift of interest,
about three orders of magnitude below what is
required to ionize the universe.

With regard to the initial mass function of Pop III stars
a new picture is emerging
from a number of recent theoretical studies of the 
collapse of primordial gas clouds at high redshifts,
induced by \h2 cooling.
At redshift $z\sim 10-30$ primordial gas clouds 
are shown to collapse to form very massive
stars with mass $M\ge 100\msun$
(Abel \etal 2002; Bromm \etal 2002; Nakamura \& Umemura 2002).
This outcome is basically determined by the Jeans
mass of the collapsing cloud, involving
a complicated interplay between cooling 
and fragmentation.
For Pop III galaxies at the redshift in question ($z\sim 13-20$)
the initial (i.e., interstellar) gas density is less
than the threshold value of $10^5~$cm$^{-3}$ 
identified by Nakamura \& Umemura (2001,2002)
and fragmentation is to
occur at low density  
with fragment mass of $\sim 100\msun$.
We adopt this new theory for Pop III stars,
called Very Massive Stars (VMS).
However, we note that the main conclusions drawn in this paper
will remain unchanged 
as long as the initial mass function of Pop III
stars is substantially top-heavy (e.g., Umeda \& Nomoto 2003).

We point out two new important sources for X-ray emission
in this section,
capable of creating a high X-ray background
at high redshift to enable the formation
of enough \h2 molecules 
to induce continuous cooling and star formation in minihalos
during the Pop III era.

\subsection{X-ray Emission from Pop III Supernova Remnants}

We use a simple analytic means to estimate the X-ray emission
from Pop III supernova remnants assuming that
a spherical supernova blast wave propagates into a uniform-density
interstellar medium.
A spherically expanding supernova remnant at sufficiently
late times follows
the simple, adiabatic, self-similar Sedov-Taylor 
solution (Sedov 1959; Taylor 1950;
Shklovsky 1968; Cox 1972; Ostriker \& Cowie 1981).
Specifically, 
the shock radius ($R_s$), 
shock velocity ($V_s$)
and postshock temperature ($T_{\rm s}$)
(for $\gamma=5/3$) obey:
\begin{equation}
R_{\rm s} = 21.7 t_4^{2/5} (E_{52}/n)^{1/5}~{\rm pc},
\end{equation}
\begin{equation}
V_{\rm s} = 83.9 t_4^{-3/5} (E_{52}/n)^{1/5}~{\rm km/s},
\end{equation}
\begin{equation}
T_{\rm s} = 8.45\times 10^6 t_4^{-6/5} (E_{52}/n)^{2/5}~{\rm K},
\end{equation}
\noindent
where $E_{52}$ is the explosion energy in $10^{52}~$ergs,
$t_4$ is the time elapsed since the onset of explosion
in $10^4~$yr and $n$ is the density of the interstellar 
medium in cm$^{-3}$.

The Sedov-Taylor phase ends with a rapid
cooling phase, resulting in the formation of a thin dense shell
(e.g., Ostriker \& McKee 1988).
This radiative cooling phase of the shocked gas
sets in abruptly approximately when 
the cooling time is equal to the time elapsed, i.e.,
when
\begin{equation}
t = {{3n_e k T_{\rm s}} \over \Lambda(T_{\rm e}) n_{\rm e}^2},  
\end{equation}
\noindent
where $100\%$ hydrogen is assumed in computing
the internal energy and the electron density $n_e$ for simplicity
(but not for the cooling function $\Lambda$; see below);
$k$ is the Boltzmann's constant;
$n_e=4n$ is the postshock electron density;
$\Lambda(T_e)$ is the volume cooling function for
a primordial plasma with $76\%$ hydrogen and
$24\%$ in helium by mass, taken from
Sutherland \& Dopita (1993).
Since the blast waves of the first generation of supernovae
propagate into a primordial gas free of metals,
metal cooling is non-existent.
More importantly, 
complex processes due to dust
need not to be considered;
in particular, X-ray emission would not suffer significantly
from dust absorption 
(Ostriker \& Silk 1973;
Burke \& Silk 1974;
Draine \& Salpeter 1979;
Shull 1980;
Wheeler, Mazurek, Sivaramakrishnan 1980;
Tielens \etal 1987;
Draine \& McKee 1993).
Combining equations (3,4) and solving for $T_{\rm s}$ 
as a function of $n$ give results 
shown in Figure 1, for two cases with $(E_{52},Z/\zsun)=(5.0,0.0)$ and 
$(0.1,1.0)$, respectively,
where the former may be appropriate for 
supernovae/hypernovae resulting from very massive Pop III stars
(Woosley \& Weaver 1982;
Ober \etal 1983;
Bond \etal 1984; 
Nakamura \etal 2001;
Heger \& Woosley 2002)
and the latter for present-day normal supernovae.

\begin{figure}
\plotone{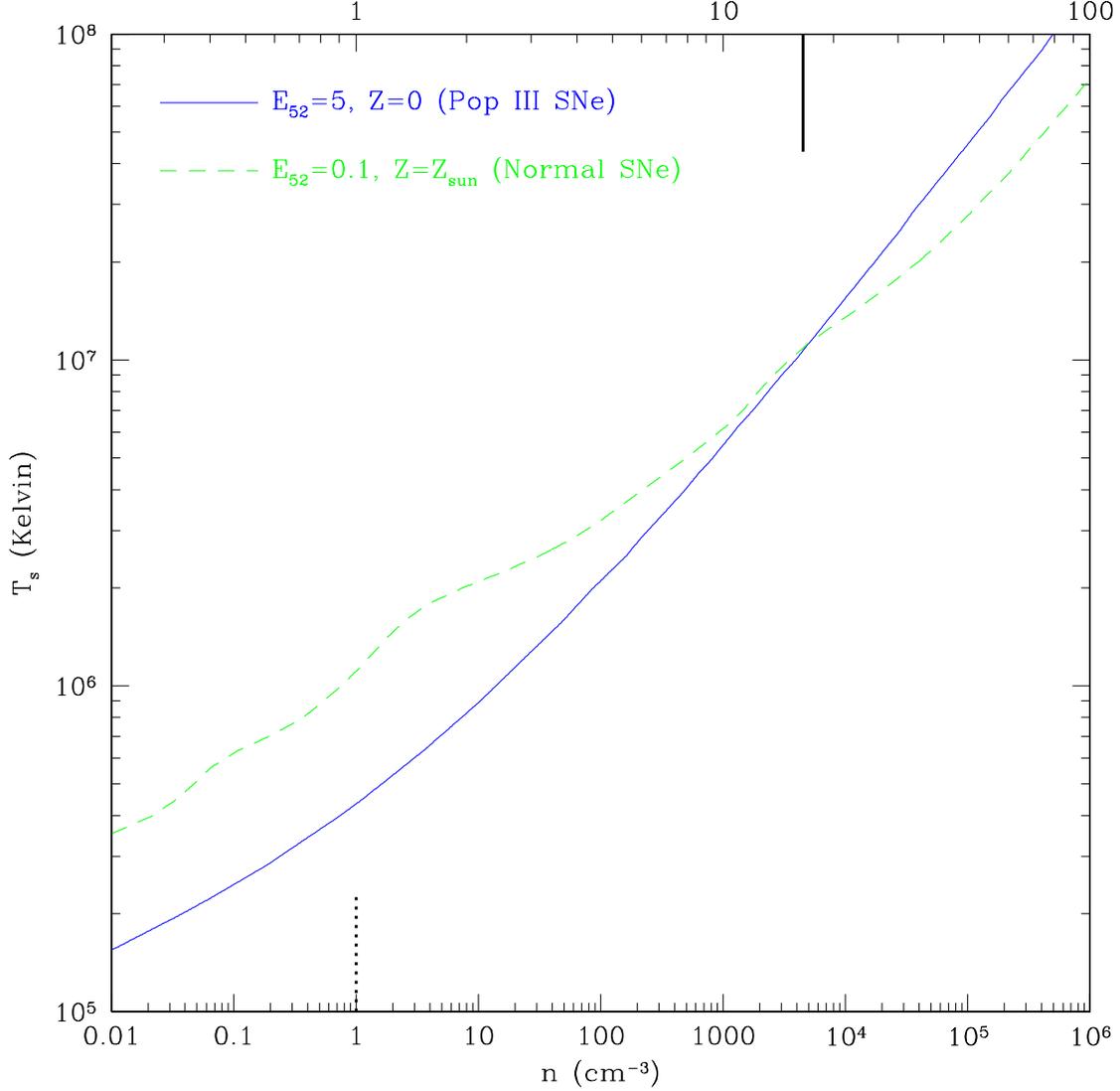}
\caption{
The solid curve shows the temperature of the cooling shock front
for a supernova/hypernova from a very massive Pop III star
with initial energy of $5\times 10^{52}~$erg
as a function of the interstellar (metal-free) gas density 
(bottom x-axis) or $1+z$ (top x-axis).
The cooling function assumes a gas of primordial composition.
The vertical tick hanging from the top x-axis
indicates redshift $z=15.5$
and the vertical tick standing on the bottom x-axis
indicates redshift $z=0$.
Cooling functions are taken from Sutherland \& Dopita (1993).
As a reference, the dashed curve is for a gas with solar metallicity
and a supernova explosion energy of $10^{51}~$erg, more appropriate
for a normal supernova.
}
\label{Ts}
\end{figure}

A critical point to note is that
the density of the general interstellar medium 
should scale with $(1+z)^3$, being significantly higher
at high redshift than that of the local interstellar medium.
This assumption should hold in the context of cosmological 
hierarchical structure formation
for the following reasons.
First, the mean gas density scales with $(1+z)^3$.
Second, halos at low and high redshift in cosmological
simulations show similar properties when
density and length are measured in their respective comoving units
(e.g., Navarro, Frenk, \& White 1997; Del Popolo 2001).
Third, the spin parameters
(i.e., angular momentum distribution) of both high and low
redshift halos have very similar
distributions 
peaking at a nearly identical value $\lambda\sim 0.05$
(Peebles 1969; White 1984; 
Barnes \& Efstathiou 1987; 
Ueda \etal 1994;
Steinmetz \& Bartelmann 1995;
Cole \& Lacey 1996; 
Bullock \etal 2001;
Cen \etal 2003).
Thus, cooling gas in galaxies at low and high redshift
should collapse by a similar factor 
before the structure becomes dynamically stable
(e.g., rotation support sets in), 
resulting in interstellar densities scaling as $(1+z)^3$.
Direct simulations (Abel \etal 2000,2002; Bromm \etal 2002)
suggest a gas density of $10^3-10^4$cm$^{-3}$ by
the end of the initial free fall for minihalos at $z\sim 20$,
verifying this simple analysis.
The same argument was given in Kaiser (1991)
on a somewhat larger scale for clusters of galaxies.
Peebles (1997), using the same scaling of density 
with respect to redshift,
suggested that an early epoch of galaxy formation may be
favored.
Observationally, the higher density in bulges 
of spiral galaxies or elliptical galaxies
are consistent with their earlier formation epochs.
In Figure 1 we have used
$n(z) = n_0 (1+z)^3$, with $n_0=1$cm$^{-3}$ being the density
of the local interstellar medium,
to translate the bottom x-axis ($n$)
to the top x-axis ($1+z$).

In Figure 1 
we see that, at $z=15.5$ (or $n=4492$cm$^{-3}$),
the blast wave
enters the rapid cooling phase 
at a postshock temperature slightly above $1.0\times 10^7$K.
Most of the energy is radiated away during the brief cooling phase
(Falle 1975,1981) with photon energy $h\nu \sim kT_{\rm s}$. 
[Subsequent evolution of the cooling shell will be subject to 
various instabilities (e.g., Chevalier \& Imamura 1982; 
Vishniac 1983; Bertschinger 1986;
Cioffi, McKee \& Bertschinger 1988).
Following that, the evolution enters 
snowplow phase driven by the 
pressure of a still hot interior gas 
(McKee \& Ostriker 1977; Ostriker \& McKee 1988).]
For a primordial gas cooling at $1.0\times 10^7$K,
it is found that $(40\%,31\%,18\%)$ of the instantaneously
radiated energy is at photon energies above
$(0.8,1.0,1.5)$keV, respectively.
Clearly, a significant amount of the total energy of the 
supernova blast wave will be turned into X-ray photons.
As we will see later, $z=15-16$ will be identified
as the redshift of the first cosmological reionization 
by Pop III stars.
At higher redshift, the emitted photons from supernova
remnant shell cooling would be still harder.

As a consistency check we find that,
for 
$n=4492$cm$^{-3}$,
at the onset of the rapid cooling phase,
the elapsed time is $t_{\rm rad}=857~$yr,
the shell radius is $r_{\rm rad}=2.09$pc,
the swept-up interstellar medium mass
is $M_{\rm rad}=4169\msun$.
The fact that $M_{\rm rad}$ is much larger
than the mass of the supernova ejecta ($M_{\rm ej}\sim 100\msun$)
guarantees the Sedov-Taylor solution
for the regime in question (until cooling sets in).
The fact that the shell radius $r_{\rm rad}$
is much smaller than the size of galaxies ($\ge 10$pc)
and $M_{\rm rad}$ is much smaller than the 
total baryonic mass in the galaxies ($\ge 10^4\msun$)
in question 
indicates that the blast wave 
at the cooling time is still sweeping through
high density interstellar medium, as was assumed.
As another consistency check,
the time required for electrons and ions
to reach temperature equilibrium 
is $t_{\rm eq}=1.4\times 10^4 E_{52}^{3/14} n^{-4/7}=162~$yr
for the case considered,
which is much shorter than $t_{\rm rad}$,
indicating that electrons can radiate away 
the shock heated thermal energy and that Sedov-Taylor phase is valid.
In contrast, in the star-burst model for AGN (Terlevich \etal 1992),
the proposed supernova remnants 
interact with a much higher density medium 
($n\sim 10^7$cm$^{-3}$)
such that the Sedov-Taylor phase is never
reached due to extremely rapid cooling 
before thermalization.

In brief, 
a large fraction of supernova explosion energy 
in Pop III galaxies at $z\sim 13-20$,
possibly as large as $f_{\rm x}\sim 0.30$,
is shown to be converted 
into X-ray photons with energy greater than $1$keV.
The X-ray background produced by this process
will be shown below to be able to 
play a positive feedback role in the formation of and cooling
by \h2 molecules in minihalos.

Let us now proceed to identify the characteristics
of the composite spectrum of the background radiation field
produced by both emissions from the VMS
and the thermal emission from VMS supernova blast waves.
A relevant parameter for our purpose is the ratio
of the energy in the LW bands ($h\nu=11.18-13.6$eV)
to the energy in photons with $h\nu > 1~$keV,
$\Psi_{\rm III} \equiv E_{\rm LW}/E_{>1keV}$.
Photons in the LW bands ($11.18-13.6$eV)
are primarily responsible for photodissociating \h2 molecules
and are not heavily absorbed by intervening intergalactic atomic hydrogen
(except for the saw-tooth modulation by the atomic Lyman line
series; Haiman, Abel, \& Rees 2000; HAR hereafter).
Hard X-ray photons at $h\nu > 1~$keV, on the other hand,
are also largely unabsorbed by intervening atomic hydrogen and helium 
and capable of penetrating deeply into the cores
of minihalos to produce free electrons through both
direct photo-ionization and secondary photo-electron ionization.
The abundance of \h2 molecules in the cores of minihalos
is primarily a result of the competition between the two.
In contrast, photons with energy in the range $13.6$eV$-1$keV 
are heavily absorbed by atomic hydrogen and helium 
prior to complete reionization of the universe and thus
have little effect on the formation of \h2 molecules in
the cores of minihalos (byt see RGS
for a positive feedback due to ionizing photons 
on the surface of the expanding HII regions and 
in relic HII regions). 
We may write $\Psi_{\rm III}$ approximately as
\begin{equation}
\Psi_{\rm III} \approx {0.007 M_{\rm VMS} c^2 f_{\rm LW} \over \xi_{\rm IMF} E_{\rm ex} f_{\rm x}},
\end{equation}
\noindent
where $f_{\rm LW}$ is the fraction of energy in
the LW bands emitted by VMS;
$c$ is the speed of light;
$M_{\rm VMS}$ is the characteristic mass of VMS ($\sim 200\msun$; see \S 3);
$E_{\rm ex}$ is the explosion energy of a typical VMS;
$\xi_{\rm IMF} < 1$ is inserted to take into account the effect
that some VMS (for example, non-rotating stars with 
$M\ge 260\msun$; Rakavy, Shaviv, \& Zinamon 1967;
Bond, Arnett, \& Carr 1984; 
Glatzel, Fricke, \& El Eid 1985;
Woosley 1986) do not produce supernovae (but collapse
wholly to black holes).
The supernova explosion energy, $E_{\rm ex}$,
released by VMS
in the mass range $140-260\msun$
is approximately $10^{52}-10^{53}~$erg
(Woosley \& Weaver 1982;
Ober \etal 1983;
Bond \etal 1984; 
Nakamura \etal 2001;
Heger \& Woosley 2002).
The VMS has approximately
a blackbody radiation spectrum with an effective temperature
of $\sim 10^{5.2}~$K (Tumlinson \& Shull 2000; Bromm \etal 2001)
for which it is found that
$f_{\rm LW}=1.5\times 10^{-2}$, yielding
\begin{equation}
\Psi_{\rm III} = 2.5 \xi_{\rm IMF}^{-1} ({f_x\over 0.3})^{-1}({E_{ex}\over 5\times 10^{52}{\rm erg}})^{-1} ({M\over 200\msun}).
\end{equation}
\noindent
Is it small enough to enhance \h2 formation in minihalos?
Let us compare to the results of a systematic
study of the effect of X-ray photons on the cooling
of minihalo gas by HAR.
HAR investigated the effect
using a $J_\nu\propto \nu^{-1}$ 
background radiation spectrum, 
with an upper cutoff at $10~$keV.
Using $\Psi$ to parameterize
their results, the findings of HAR are that, when
\begin{equation}
\Psi_{\rm HAR} \le {\ln (13.6/11.18)\over \epsilon_x \ln (10/1)} = 5.1
\end{equation}
\noindent
(for $\epsilon_x=0.016$),
\h2 formation is enhanced and cooling by \h2 is sufficient to allow for 
the gas in the cores of 
minihalos with virial temperature $T_{\rm v}\ge 1000$K
to collapse to form stars.
We note that in evaluating $\Psi_{\rm HAR}$ above
we have taken into account a factor of $\sim 6$ 
underestimation of $\Psi_{\rm HAR}$ in the original 
work of HAR due to 
a factor of $6$ underestimation of the 
LW photon absorption cross section by \h2 molecules
(Haiman 2002, private communications; RGS;
Glover \& Brand 2002).
The correction factor $6$ is a lower limit in the 
optically thin case for \h2 molecules for the
cores of minihalos.
In the optically thick limit, a substantially larger
correction factor needs to be applied 
(i.e., $\Psi_{\rm HAR}$ would be substantially larger than
$5.1$), although
the exact effect would require a re-calculation
of the results in HAR.
It thus appears that 
the X-ray emission from Pop III supernova remnants alone
may be able to produce enough X-ray photons relative to the number of
photons in the LW bands such that production rate of 
\h2 molecules dominates over the destruction rate
in most of the minihalos.

\subsection{X-ray Emission from Miniquasars Powered by Pop III Black Holes}

Another direct 
consequence of Pop III VMS
is an inevitable production of a significant amount of black holes.
Pop III VMS more massive than $260\msun$ would eventually implode 
carrying the entire mass to form black holes 
(Rakavy, Shaviv, \& Zinamon 1967; Bond, Arnett, \& Carr 1984; 
Glatzel, Fricke, \& El Eid 1985;
Woosley 1986) without producing explosions.
Pop III VMS less massive than $140\msun$,
after about three million years of luminous life,
would explode as supernovae or hypernovae leaving behind black holes
of mass $\sim 10-50\msun$ (Heger \& Woosley 2002).
These Pop III black holes could accrete gas from
surroundings to shine as miniquasars.
The likelihood of gas accreting onto these black holes
is at least as high as in their lower redshift counterparts 
since the gas density is higher and halos
tend to have somewhat lower spins at high redshift.
In the case of the very massive Pop III black holes ($M\ge 260\msun$)
formed without explosion, gas may be ready to accrete immediately,
as the surrounding gas has not been blown away.
The dynamics of the black holes in these environments
in the context of cosmological structure formation 
is a complex subject and can only be treated
in separate works. 
Here we make the assumption that these black holes
would accrete gas and grow.

The characteristic gas temperature of a disk
powered by accretion onto a black hole
is (Rees 1984):
\begin{equation}
T_{\rm E} = 1.3\times 10^7 M_2^{-1/4}~K,
\end{equation}
\noindent
where $M_2$ is the black mass in $10^2\msun$.
For $M_{\rm BH}=(200,20)\msun$, we have 
$T_{\rm E} = (1.3\times 10^7, 2.4\times 10^7)~$K.
While the spectral energy distribution (SED) of quasars 
powered by supermassive black holes 
is known to contain a significant fraction 
in X-rays (e.g., Elvis \etal 1994),
the miniquasars at high redshift powered
by much smaller black holes 
will be conspicuous in X-rays, probably emitting
predominantly in X-ray band from both thermal
and nonthermal emission.
A somewhat more quantitative argument may be made
as follows.
The SED of observed quasars powered by supermassive black holes
of mass $\sim 10^8\msun$ contains
a substantial amount of energy in the X-rays
but the largest concentration of energy appears to peak
at $\sim 12-13$eV (UV bump; Elvis \etal 1994) barring
the unknown gap between UV and X-ray.
Under the reasonable assumption that the SED peak frequency 
scales with the characteristic temperature $T_{\rm E}$,
then the peak frequency for Pop III black hole powered miniquasars
would be shifted (by a factor of $\sim 100$ compared to quasars)
to $h\nu \sim 1$keV.

Let us now compute the X-ray background 
produced by miniquasars powered by Pop III black holes.
We put this quantitatively in the context of
the ratio of energy in LW photons to that of X-ray photons.
First, an order-of-magnitude estimate.
If a miniquasar has accreted
a fraction, $\eta_Q$, of the initial mass of the black hole,
the radiated energy will be
\begin{equation}
E_Q = \eta_Q \alpha c^2 M_{BH},
\end{equation}
\noindent
where $\alpha$ is the radiative efficiency.
Then, the ratio $\Psi$ may be written as 
\begin{equation}
\Psi_Q = {0.007 c^2 f_{\rm III} f_{\rm LW} \over 0.1 \alpha_{0.1} c^2 \eta_Q f_{\rm III}f_{\rm BH} f_{\rm Xray}f_{\rm esq}} = 10^{-3} \eta_Q^{-1} \alpha_{0.1}^{-1} f_{BH}^{-1} f_{\rm Xray}^{-1}f_{\rm esq} ,
\end{equation}
\noindent
where $f_{\rm III}$ is the fraction of mass collapsed into
Pop III stars; 
$f_{\rm BH}$ is the fraction of mass in $f_{\rm III}$ that
ends in black holes;
$\alpha_{0.1}$ is radiative efficiency of the miniquasars in units
of $0.1$;
$f_{\rm Xray}$ is the fraction of energy radiated by miniquasars
with photon energy greater than $1~$keV;
$f_{\rm esq}$ is the X-ray escape fraction from miniquasars;
we have used $f_{\rm LW}=1.5\times 10^{-2}$. 
The value of $\alpha_{0.1}$ is thought to be close to unity (Rees 1984)
and the value of $f_{\rm Xray}$ is expected to be of order unity.
The fraction of mass in Pop III stars ultimately collapsing to black holes
$f_{\rm BH}$ is also of order unity, if either Pop III stars 
are very massive $\sim 100\msun$ or $\gg 1\msun$.
The accretion fraction $\eta_Q$ would depend on the gas
accretion rate and the formation time of the black hole;
parameterization by Equation (10) would be fairly accurate
if black holes formed long time ago.
But a more accurate way to compute the X-ray energy from
miniquasars is to integrate over time over all 
miniquasars by parameterizing each radiating at
$\xi_{\rm Edd}$ times the Eddington luminosity.
The results are shown in Figure 2,
where the solid and dotted curves show $\Psi_{\rm Q}$ 
with radiative efficiency of $\alpha=(0.1,0.03)$, respectively.
The global Pop III star formation rate 
is computed using Press-Schechter (1974) formula 
considering all halos with virial temperature greater than
$1000~$K. 
The reason that the dotted curve 
with lower radiative efficiency ($\alpha=0.03$)
lie below (i.e., emits more X-ray photons than)
the solid curve is that the black hole mass in this case
grows at a faster rate, being inversely proportional to the radiative
efficiency.

\begin{figure}
\plotone{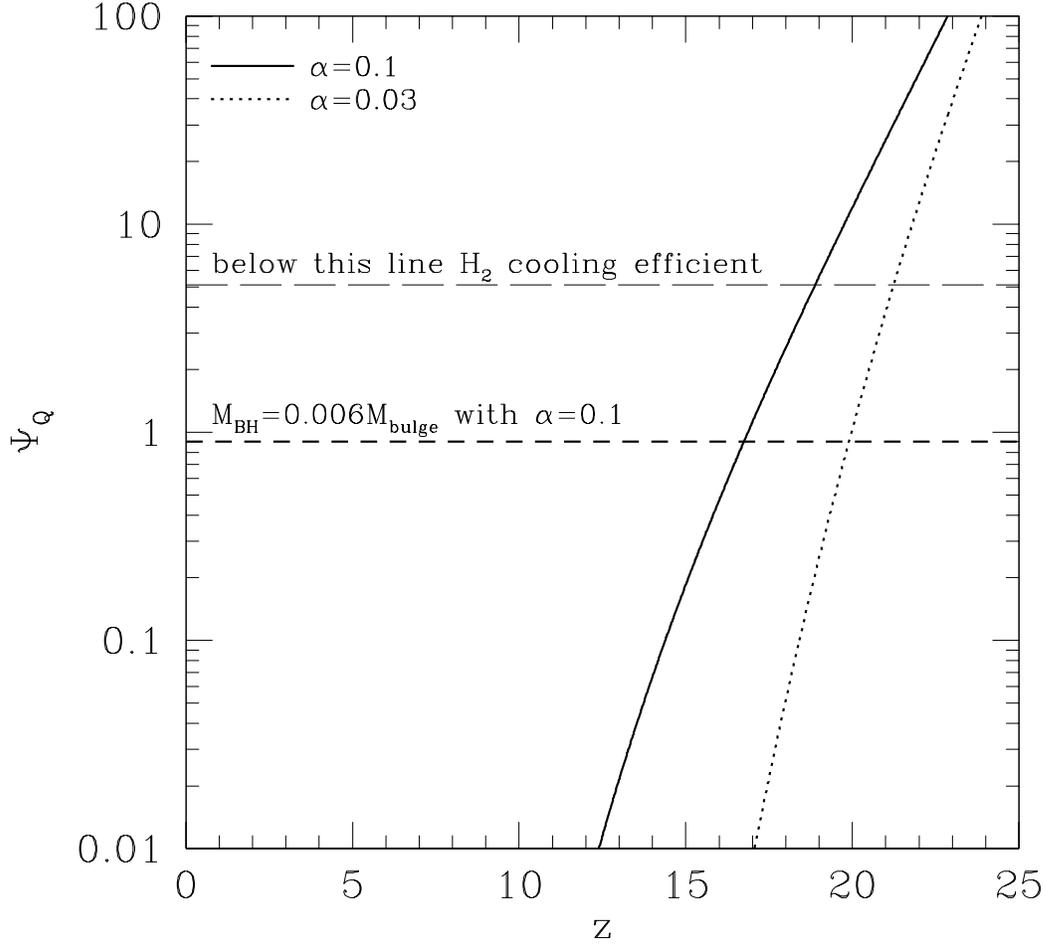}
\caption{
The solid and dotted curves show $\Psi_{\rm Q}$ as a function of redshift,
by integrating over time the miniquasar luminosity 
radiating at Eddington luminosity
with radiative efficiency of $\alpha=(0.1,0.03)$, respectively.
The horizontal dashed line indicates
the case if one adopts the local observed 
black hole mass-bulge mass relation (Magorrian \etal 1997)
with radiative efficiency of $\alpha=0.1$.
The horizontal long-dashed line indicates the required value (HAR),
below which \h2 cooling is efficient.
$\Psi_{\rm Q}$ would scale with $\xi_{\rm Edd}$
and $f_{\rm BH}^{-1} f_{\rm Xray}^{-1}$, for both of which
we have assumed value of unity.
We also assume that $20\%$ ($f_{\rm esq}=0.2$)
of the soft X-ray radiation ($\sim 1~$keV),
averaged over all miniquasars, escapes to the intergalactic space; 
this value is based on observations of AGN and quasars at lower redshift
(Brandt 2002, private communications).
We have ignored cosmological redshift and
dilution effects in this illustration.
}
\label{QSO}
\end{figure}

Another useful way to estimate
the growth of mass in black holes 
is to follow the local observed 
black hole mass-bulge mass relation,
$M_{\rm BH}\sim 0.006 M_{\rm bulge}$,
(Magorrian \etal 1998).
If the same ratio of the mass in black holes to bulge mass,
which at high redshift may be equated to the total stellar
mass, holds at high redshift,
then we can readily compute the energy liberated
by gas accretion onto Pop III black holes as
\begin{equation}
\Psi_{\rm Q} = {0.007 c^2 f_{\rm III} f_{\rm LW} \over 0.006\times 0.1 \alpha_{0.1} c^2 f_{\rm III}f_{\rm BH} f_{\rm Xray}f_{\rm esq}} = 0.88 \alpha_{0.1}^{-1} \left({f_{\rm esq}\over 0.2}\right)f_{BH}^{-1} f_{\rm Xray}^{-1}.
\end{equation}
\noindent
The horizontal dashed line in Figure 2 indicates
$\Psi_{\rm Q}$ computed this way
with radiative efficiency of $\alpha=0.1$ and
X-ray escape fraction $f_{\rm esq}=0.2$.

From Figure 2 it is evident that $\Psi_Q$ is significantly
smaller than $\Psi_{HAR}=5.1$ (Equation 7), 
at $z\le 20$, required to induce sufficient \h2 cooling in minihalos (HAR).
The higher value of $\Psi_{\rm Q}$ 
at higher redshift ($z\ge 20$)
is due to the fact that the age of the universe
becomes much shorter than the Eddington time of $\sim 4\times 10^8~$yr
and the black holes have not had enough time to 
accrete a substantial amount of gas;
the age of the universe is $1.9\times 10^8~$yr
at $z=20$ (for $H=70~$km/s/Mpc, $\Omega_M=0.25$, $\Lambda=0.75$).

In summary,
two distinctive mechanisms, namely
supernova remnants and miniquasars,
each appears to be able to 
generate enough X-ray emission at $z\sim 13-20$
to provide positive feedback on star formation in subsequent
(other) minihalos by making a sufficient number of X-ray photons
relative to the number of destructive photons produced
in the LW bands.
The combination of the two should ensure that
enough X-ray radiation is produced as a result
of Pop III star formation.
In addition, there may be other, significant
X-ray emission mechanisms such as inverse
Compton emission (e.g., Hogan \& Layzer 1979; Oh 2001) 
or massive X-ray binaries (e.g., Bookbinder \etal 1980;
Helfand \& Moran 2001).
Moreover, additional positive feedback mechanisms,
such as that proposed by Ferrara (1998) due to
enhanced \h2 formation in the supernova cooling shells,
that put forth by RGS
from the enhanced \h2 formation at the surfaces
of Str\"omgren
spheres of individual Pop III galaxies
and that quantified in the next section (\S 3) due to
\h2 formation in relic H II regions produced by
Pop III galaxies 
will further help promote \h2 formation.
Taking all the processes together, it appears
highly likely that 
the chief obstacle to continuous \h2 formation 
and cooling in minihalos is removed.

Since the assumption that X-ray emission produced 
by these two processes related to Pop III star formation 
can produce positive feedback to subsequent
star formation is important,
it is warranted to have independent checks.
We will compare our results to Glover \& Brand  (2002; GB hereafter).
We compare to their X-ray emission model due to
inverse Compton, which has luminosity density
$L_{\rm X} = 7.7\times 10^{23} (\nu_0/\nu) f_{\rm e}~{\rm erg}~{\rm s}^{-1} {\rm Hz}^{-1} (\msun {\rm yr}^{-1})^{-1}$,
where $\nu_0=1$keV and $f_{\rm e}$ is the fraction of 
supernova energy transferred
to relativistic electrons.
Integrating 
$L_{\rm X}$ over a range $1-10$keV 
we obtain
$E_{\rm X} = 4.28\times 10^{41} f_{\rm e}~{\rm erg}~{\rm s}^{-1} (\msun {\rm yr}^{-1})^{-1}$.
The luminosity at the LW bands in GB
is 
$L_{\rm LW} = 1.1\times 10^{28} {\rm erg}~{\rm s}^{-1}~{\rm Hz}^{-1} (\msun {\rm yr}^{-1})^{-1}$, which after integration over LW bands gives
$E_{\rm LW} = 6.4\times 10^{42} {\rm erg}~{\rm s}^{-1}~(\msun {\rm yr}^{-1})^{-1}$.
Taking the ratio of the above two luminosities yields
$\Psi_{\rm GB} \equiv E_{\rm LW}/E_{\rm X} = 150$,
for the most optimistic model that they use with
$f_{\rm e}=0.1$.
For this model GB find that
the critical virial temperature of halos where cooling is sufficient
is lowered from $8\times 10^3$K to about $(2-3)\times 10^3$K
at $z\sim 17$ (see Figure 
6 in GB), a significant effect.
While even this model is already sufficient to enable 
molecular cooling in halos with $T_{v}>2000$K to reionize
the universe at $z\sim 17$ (see Figure 5 below),
it is expected that the much smaller $\Psi_{\rm III}$
computed above for Pop III stars in our model would further drive 
the critical virial temperature to much 
lower values and enable star formation in almost all
minihalos,
consistent with the conclusions reached by HAR.

\section{Formation of \h2 Molecules in Relic HII Regions Produced by Pop III Galaxies} 

While the positive feedback due to 
a high X-ray background produced by Pop III supernovae
and miniquasars powered by Pop III black holes
seems sufficient to sustain continuous gas cooling 
in gaseous minihalos, we quantify another important source
for production of \h2,
which was first pointed out by RGS.

Each Pop III galaxy creates an HII region (Str\"omgren sphere) around it.
Since the lifetime of a Pop III galaxy
of $3\times 10^6~$yr is comparable
to and somewhat shorter than the 
hydrogen recombination time, 
$t_{\rm rec}= 1.3\times 10^7$yr at $z\sim 20$,
we may assume for simplicity, without causing substantial errors,
that the size of the Str\"omgren 
sphere is just determined by the number of 
ionizing photons emitted, $N_{\rm ion}$.
We follow the evolution of such a relic HII region
after the death of a Pop III galaxy, in the absence
of any external radiation field.
We solve a set of rate equations
and energy equation for relevant species in
an H II region free of metals.
The relevant reaction coefficients 
are taken from Abel \etal (1997).

\begin{figure}
\plotone{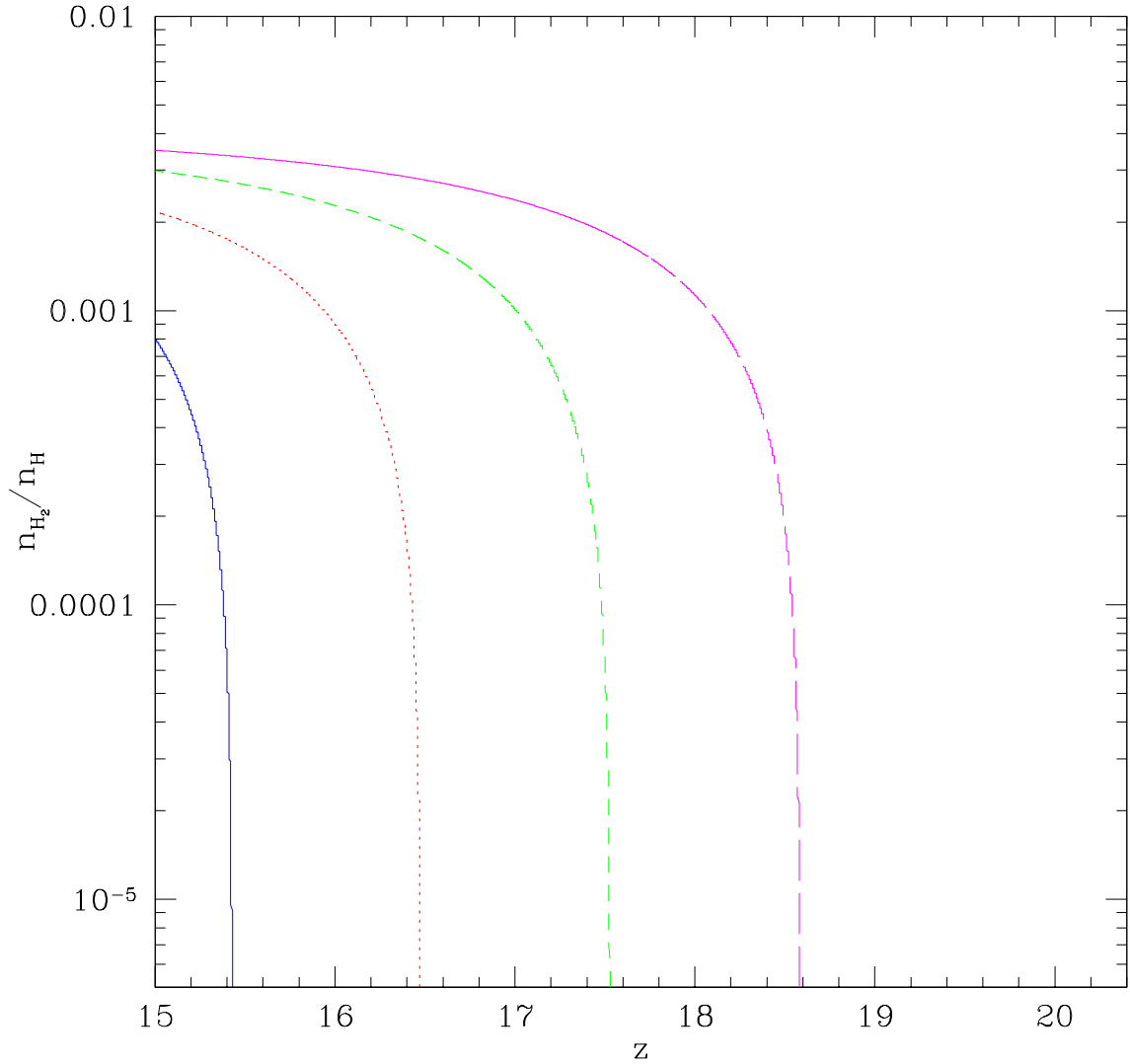}
\caption{
shows the fraction of \h2 molecules
as a function of redshift for four
relic HII regions formed at $z=(17,18,19,20)$,
respectively.
We assume a reasonable and perhaps conservative 
clumping factor of $15$ for HII regions 
in the vicinity of Pop III galaxies.
}
\label{h2}
\end{figure}

\begin{figure}
\plotone{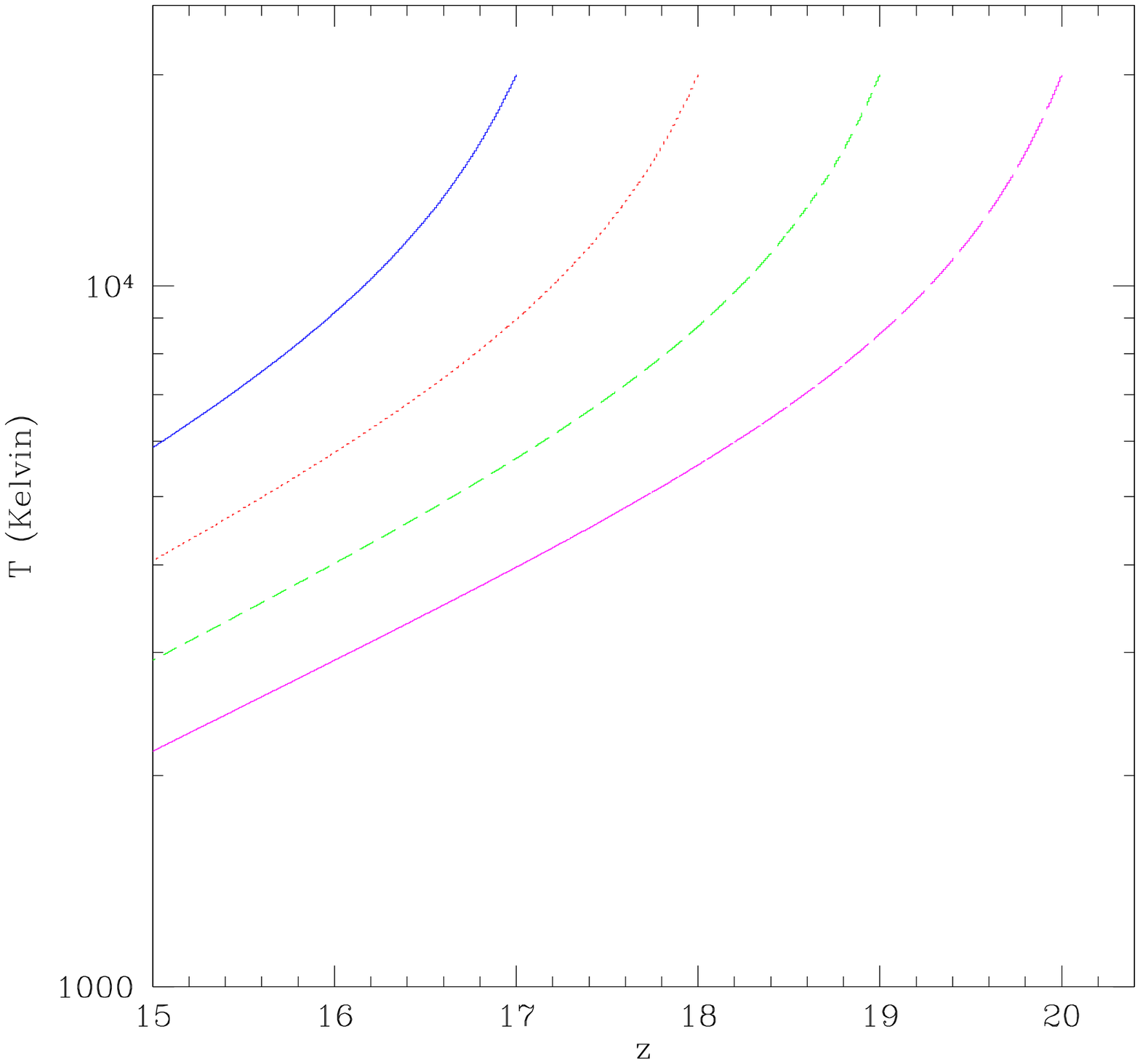}
\caption{
shows the evolution of temperature 
for four H II regions formed at $z=(17,18,19,20)$,
respectively.
}
\label{h2T}
\end{figure}

Figure 3 shows the fraction of \h2 molecules
as a function of redshift for four
H II regions formed at $z=(17,18,19,20)$,
evolved up to redshift $z=15$, which will be identified as
the first reionization epoch by Pop III stars (\S 4).
We see that the \h2 fraction in H II regions 
shows a sharp rise at an early time 
and flattens out at later times.
The reason for this asymtotic convergence may be understood 
by examining Figure 4, where
the temperature evolution of the gas in the region
is shown.
\h2 formation rate decreases with 
decreasing temperature (Abel \etal 1997), 
causing the total \h2 abundance to flatten out below $1000~$K. 
On the high temperature end,
since \h2 formation rate only becomes 
significant after the temperature has dropped below about $7000~$K,
we see a slight delay ($\Delta z\sim 1.5$) in the ascent of the 
\h2 abundance in Figure 3.

Let us now assess the significance of the
\h2 molecules formed in these relic H II regions.
A critical ratio here 
is the number of \h2 molecules produced 
over the number of LW photons produced,
denoted as $R_{\rm relic}$.
For $N_{\rm ion}$ ionizing photons emitted by Pop III stars,
there will be approximately $N_{\rm HII}=N_{\rm ion}$ 
hydrogen atoms ionized in the H II region.
At the same time, the number of LW photons
produced by the same stars 
is $1.5\times 10^{-2}N_{\rm ion}$.
From Figure 3, we take an approximate average
$n_{H_2}/n_H=2\times 10^{-3}$ for
the abundance of \h2 molecules produced by $z=15$.
Then, we obtain $R_{\rm relic}=0.13$.
Since about $10$ LW photons 
are needed to photodissociate one \h2 molecule
(RGS; Glover \& Bland 2002),
the break-even point for the ratio
of the number of \h2 molecules produced
over the number of LW photons produced 
is about $0.1$,
i.e., no net molecules would be produced.
It thus appears that \h2 molecules produced
in the relic H II regions produced by Pop III stars
may be sufficient to offset 
the production of LW photons by the same stars.
Consequently, LW radiation background may never
be able to build up during the Pop III era.
However, a more careful calculation,
taking into account the density fluctuations among others,
is needed to precisely quantify this positive feedback effect.
But we defer such a calculation to a future work.
Nevertheless,
it seems possible that this positive feedback in relic H II
regions may be more effective than 
that proposed by Ferrara (1998) due to
enhanced \h2 formation in the supernova cooling shells
and that put forth by RGS
from the enhanced \h2 formation at the surfaces
of Str\"omgren spheres of individual Pop III galaxies.

In summary, it appears that LW radiation background 
may have never been able to build up to destroy
\h2 molecules in subsequent minihalos.
In the event that LW radiation background was able to
build up,
its intensity would have been substantially reduced.
The fact that the X-ray background produced by Pop III
supernovae and miniquasars powered by Pop III black holes
have already been capable of countering
an undiminished LW radiation background (i.e.,
by ignoring all the \h2 molecules produced by 
these positive feedback processes) leads us to conclude that
subsequent star formation in minihalos will not be 
hindered by previous star formation in the Pop III era.

\section{The Process of Cosmological Reionization}

\subsection{Computational Method}

We will examine the evolutionary history 
of the IGM from a very early redshift ($z\gg 20$)
until the universe is completely reionized 
at $z\sim 6$, as suggested by recent
observations. 
Rather than taking a brute-force approach,
which will be deferred to a later work,
we will use a new, fast Monte Carlo-like approach to 
economically sample parameter space.
This method, we hope,
should be capable of capturing the essential physical processes involved.

At any time the gas in the universe
consists of $N$ distinct regions
in the two-dimensional
phase space specified by $(T_i,y_i)$, where $T_i$ and $y_i$ are
the temperature and neutral hydrogen fraction, respectively,
and $N$ is a varying number.
Each phase space region $i$ 
contains $f_i$ mass fraction of the total gas in the universe
and its mean gas density is assumed to be equal
to the global mean; the sum of all $f_i$ is unity.
For each region $i$ we solve a combined set of equations simultaneously 
to follow the coupled evolution of gas and star formation:
\begin{equation}
{df_i \over dt} = -{dN_{ion}\over dt} {1\over \bar n\bar y} f_i,
\end{equation}
\noindent
\begin{equation}
{dT_i \over dt} = -{\Sigma_i(T_i,y_i,C_i,z) \over 3k},
\end{equation}
\noindent
\begin{equation}
{dy_i \over dt} = \alpha C_i \bar n (1-y_i)^2 - \beta C_i \bar n y_i(1-y_i),
\end{equation}
\noindent
where
$\bar n$ is the global mean hydrogen number density;
$\bar y$ is the global average of the neutral hydrogen fraction;
$\Sigma_i$ is the net cooling rate per unit mass
(including Compton cooling, atomic line cooling, recombination cooling,
photoheating due to ionization and adiabatic cooling due to 
cosmic expansion),
which is a function of $T_i$, $y_i$, clumping factor $C_i$ (see below)
and redshift $z$;
$k$ is the Boltzmann's constant;
$\alpha$ is the case B hydrogen recombination coefficient;
$\beta$ is the hydrogen collisional ionization coefficient;
$dN_{ion}/dt$ is the number of ionizing photons emitted per unit volume
per unit time,
equal to 
\begin{equation}
{dN_{ion} \over dt} = {c^* \epsilon_{UV} c^2 \bar\rho f_{\rm es}\over 13.6{\rm eV}} {d\psi\over dt}
\end{equation}
\noindent
where 
$c^*$ is the star formation efficiency;
$\epsilon_{UV}$ is the fraction of energy (in units of rest mass of
stars formed) turned into hydrogen ionizing photons;
$\bar\rho$ is the global mean gas mass density;
$c$ is the speed of light;
$f_{\rm es}$ is the ionizing photon escape fraction from galaxies;
${d\psi\over dt}$ is the global rate of 
fraction of gas formed into stars,
equal to 
\begin{equation}
{d\psi \over dt} = \sum_{i=1}^N f_i {d\psi_i\over dt},
\end{equation}
\noindent
where ${d\psi_i\over dt}$ is the rate of the fraction of gas formed
into stars in region $i$.
At each time step,
we use the Press-Schechter (1974) formula:
\begin{equation}
\psi_i(>M,z) = {\rm erfc}({\delta_c\over \sqrt{2}\sigma(M,z)}) 
\end{equation}
\noindent
to compute 
the fraction of gas, $\psi_i$, that is
in halos in region $i$;
where ${\rm erfc}()$ is the complimentary error function 
and $\sigma(M,z)$ is the rms density fluctuation
smoothed over a region of mass $M$ at $z$;
$\delta_c$ is a constant equal to $1.69$ indicating
the amplitude of the density fluctuations 
on a top-hat sphere that collapses at $z$ 
(Gunn \& Gott 1972). 
The threshold halo mass $M$ is basically the Jeans mass,
given the IGM temperature of region $i$;
i.e., only those halos whose mass is greater than
the Jeans mass at that time
will accrete gas
and
contribute to the star formation at that time step.
In practice, however, the threshold mass $M$ depends on
the history of the gas involved as well as non-trivial 
effects due to dynamically dominant dark matter
and is thus best determined by detailed simulations.
We will use the empirically determined ``filter mass" formula
from full hydrodynamic simulations by Gnedin (2000b):
\begin{equation}
M_{f}= 1.0\times 10^{8} \kappa ({\Omega_M\over 0.25})^{-1/2} \left({T_i\over 10^4{\rm K}}\right)^{3/2} \left({1+z\over 10}\right)^{-1.5},
\end{equation}
\noindent
where at the relevant redshift range, $z\sim 6-20$,
the empirically determined constant 
$\kappa$ is found to be $0.5-1.6$.
The results obtained in \S 4.2,4.3
do not depend on 
$\kappa$ sensitively in the indicated range of $\kappa$.

As has been validated in \S 2,3 at the Pop III era,
star formation occurs in all halos where gas is able to collect, 
not limited to large halos where atomic hydrogen cooling becomes efficient.
It is, however,
unclear whether \h2 cooling would still be efficient
in the Pop II era.
We show that even in the absence of the X-ray background, 
a high residual ionization fraction can be maintained
after the first reionization,
thanks to the long hydrogen recombination time.
Let us make a simple estimate 
by computing the ratio of the recombination
time to the dynamical time of a region of overdensity $\delta$
with ionization fraction $x$.
We find 
\begin{equation}
{t_{\rm rec}\over t_{\rm dyn}}=0.8 ({1+z\over 16})^{-3/2} ({\delta\over 10^5})^{-1/2} ({x\over 10^{-2}})^{-1}
\end{equation}
\noindent
for $T=5000~$K.
It is thus clear that 
an ionization fraction of order $10^{-3}-10^{-2}$ will be maintained
for $\delta$ as high as $10^5$ (the core density is about
$\delta=10^4-10^5$),
as long as the gas is ionized at least to that level at a higher
redshift (of course, the first reionization raises
the gas ionization to a much higher level than this). 
Such a level of ionization is adequate to enable the formation of 
a sufficient amount of \h2 molecules
to supply ample cooling at $T\le 10^4$K (HAR; Tegmark \etal 1997)
and 
star formation should continue to take place in any gaseous halo.
Looking forward to the results,
it turns out that the condition for \h2 cooling
has a rather small effect for the second cosmological reionization,
since the amount of intergalactic gas able to 
accrete onto minihalos is very small.

The halo mass within the virial radius
$M_{\rm v}$ is related to the virial temperature
at $z$ by the standard formula:
\begin{equation}
M_{\rm v} = 1.2\times 10^8 \left({\Omega_{\rm M}\over 0.25}\right)^{1/2} \left({1+z\over 10}\right)^{-3/2} \left({T_{\rm v}\over 10^4{\rm K}}\right)^{3/2} \msun,
\end{equation}
\noindent
where $T_{\rm v}$ is the virial temperature in Kelvin
and $\Omega_{\rm M}$ is the matter density at zero redshift
in units of the closure density.
Figure 5 relates the virial temperature at each redshift
to the halo mass for the five indicated virial temperatures.
Figure 6 shows the fraction of gas in halos
as a function of redshift,
with virial temperatures greater than
the five indicated values,
$T_v=(200,1000,2000,4000,8000)$K.

\begin{figure}
\plotone{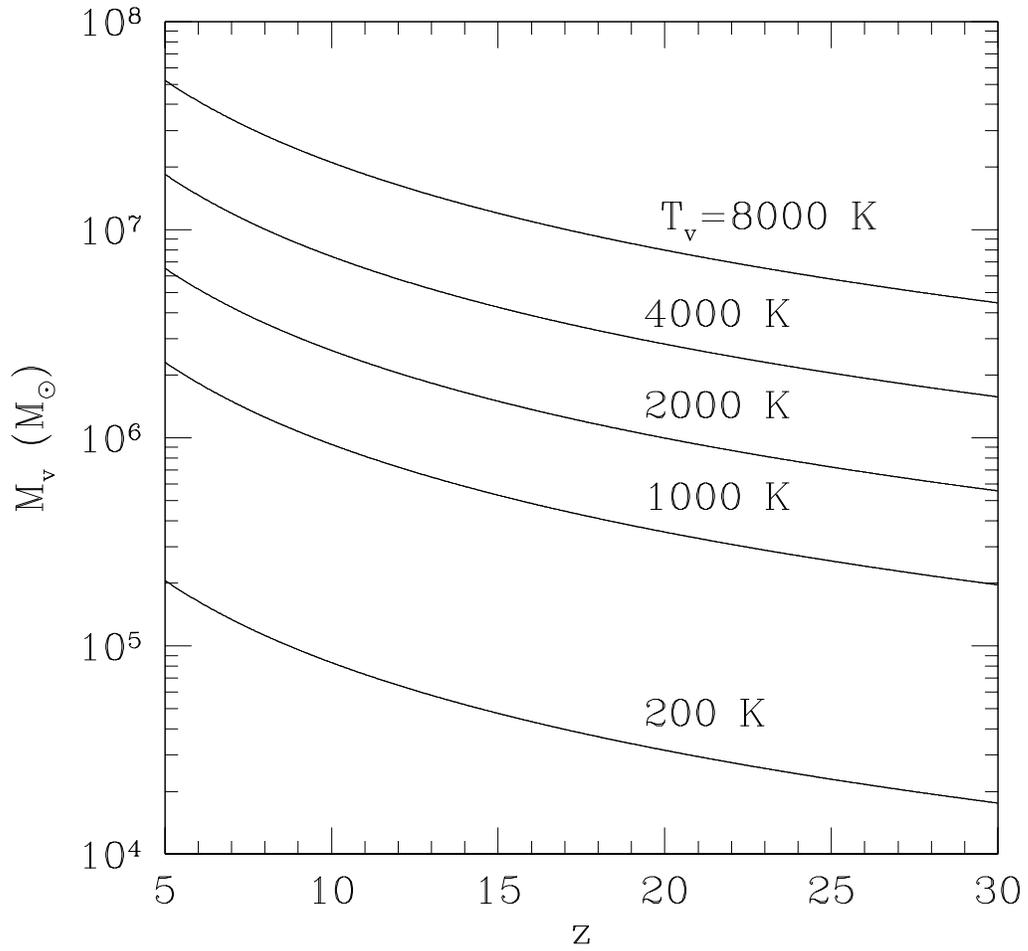}
\caption{
relates the virial temperature to the mass
of the halo as a function of redshift,
for five indicated virial temperatures,
$T_v=(200,1000,2000,4000,8000)$K.
}
\label{MZ}
\end{figure}

\begin{figure}
\plotone{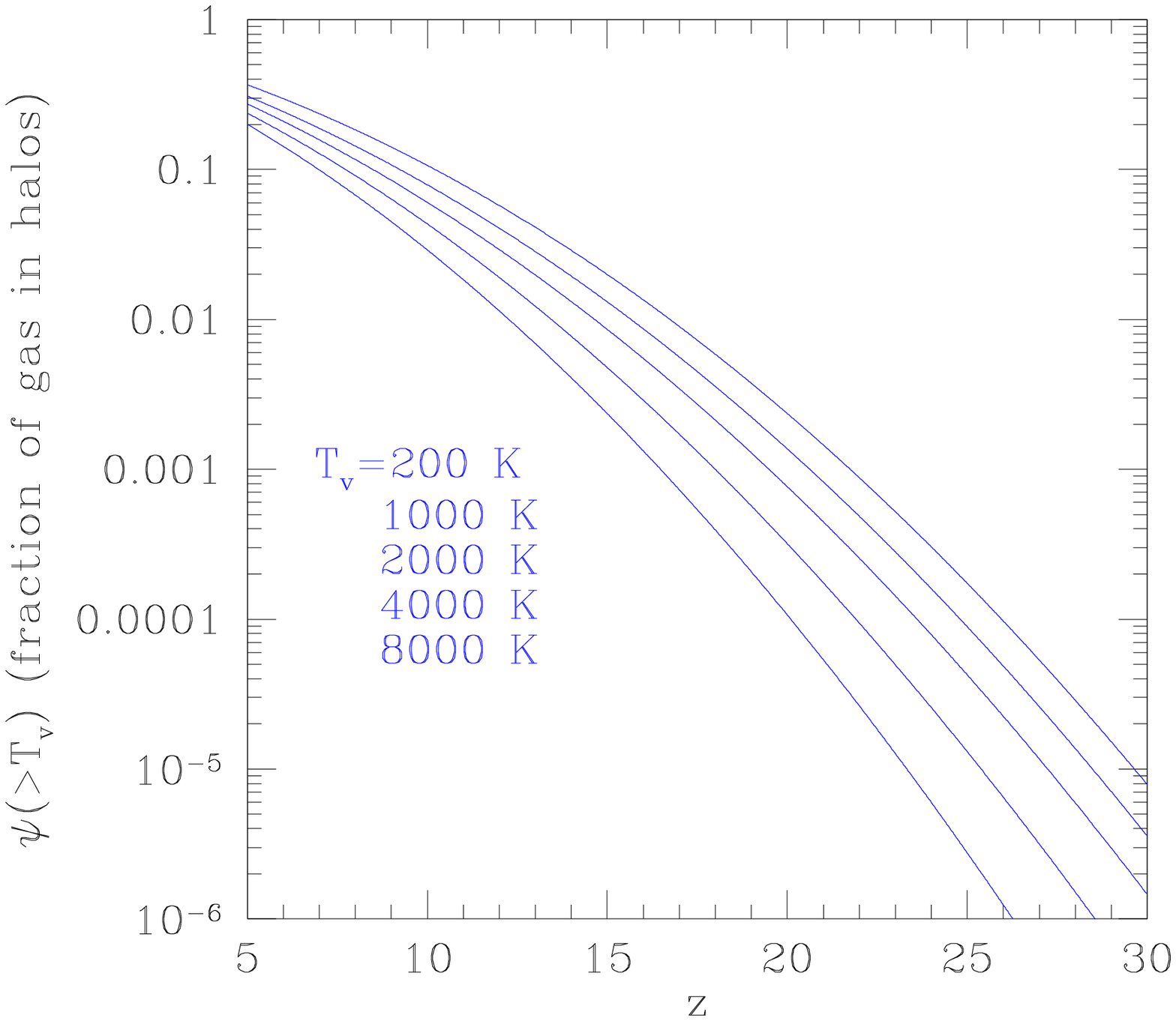}
\caption{
shows the fraction of gas in halos
as a function of redshift,
for five indicated virial temperatures,
$T_v=(200,1000,2000,4000,8000)$K.
}
\label{Ztemp}
\end{figure}

The clumping factor in each region affects
local atomic processes and is defined to be
$C_i=\psi_i C_{\rm halo} + (1-\psi_i)$,
where $C_{\rm halo}$ is the effective clumping factor 
of virialized gas and the remaining gas is 
assumed to have a clumping factor equal to unity for simplicity.
We will adjust $C_{\rm halo}$ such that the overall
$<C>$ averaged over all regions matches detailed simulations
at some appropriate redshift.
Note that the adjustment on $C_{\rm halo}$ is necessary
in order to take into account complicated radiation shielding effect
affecting halos.

Having defined the relevant formulae 
let us go back to examine the main evolution equations more closely.
Equation (12) is the rate equation that mass in region $i$
is photoionized;
i.e., the rate of the amount of gas swept by the ionization front.
The sum of ionized gas over all the phase space regions
at each time step
creates a separate (HII) region in the phase space with
a mean temperature equal to 
\begin{equation}
T_{N+1}= \sum_{i=1}^N f_i (T_i + y_i T_{ion}),
\end{equation}
\noindent
where the first term inside the parentheses on the right hand side
is the temperature of region $i$ and the second term
accounts for photo-heating,
with $T_{ion}$ set to $2\times 10^4$K 
(varying $T_{ion}$ from $1\times 10^4$K to
$3\times 10^4$K does not significantly change the results)
We set the initial neutral fraction of the newly created
HII region to $y_{N+1}=10^{-4}~$,
although results obtained do not depend on the exact choice for this value,
as long as it is a small number.
The initial mass fraction of this new HII region is equal to 
\begin{equation}
f_{N+1} = {dN_{ion}\over dt} {1\over \bar n\bar y} \Delta t,
\end{equation}
\noindent
where $\Delta t$ is the time step.
We formulate that all phase space regions 
co-exist in real space 
{\it when averaged over a sufficiently
large volume} and photoionization ionizes
all phase space regions proportionally,
as indicated by Equation (12).
Figures (13,14) are the energy equation 
and rate equation for neutral hydrogen 
in each phase space region.

The attentive reader may have noticed the central idea
behind this simple approach.
That is, this approach attempts to mimick the stochastic
percolation process during cosmological reionization,
where individual HII regions
are created and their subsequent evolution followed.
New HII regions created may originate from a combination of 
regions, some of which have previously been ionized but subsequently
recombined, some of which were never ionized
and some of which may be recently created HII regions but now
get percolated.
We do not include a photoionization term
on the right hand side of Equation (14),
rather, equivalently,
we remove the fraction of gas in that region that is
being ionized  and add it to a new phase space region
according to Equations (12 and 22).
This is equivalent to saying that 
gas is ionized when ionization front passes through,
and is not ionized uniformly.

One important difference between this method and 
another commonly used method 
which computes the filling factor of HII regions 
(e.g., Shapiro \& Giroux 1987; Haiman \& Loeb 1997;
Madau \etal 1999; Miralda-Escud\'e, Haehnelt, \& Rees 2000)
is that we follow the evolution of all regions
including HI, HII regions and partially ionized regions.
This is important for our purpose,
chiefly because at high redshift HII regions 
experience rapid recombination and cooling.
Needless to say, this method is still highly simplified.
More sophisticated calculations would require 
simulations that follow detailed three-dimensional
radiative transfer (e.g.,
Norman, Paschos, \& Abel 1998;
Abel, Norman, \& Madau 1999;
Razoumov \& Scott 1999;
Kessel-Deynet \& Burkert 2000;
Gnedin \& Abel 2001; 
Cen 2002)
but will be carried out in our future work.

We start a simulation at a very early redshift, say, $z=100$,
when the gas may be safely assumed to cold and neutral.
For simplicity we set the initial gas temperature
to be equal to the microwave temperature,
although the gas may be cooler than that,
but the results would not change noticeably.
Thus, the initial condition is this:
$T_1=2.73 (1+z_i)$K (where $z_i$ is the starting redshift),
$y_1=1$, 
$C_1=1$, $f_1=1$ 
and the initial number of entries in the phase space
$(T,y)$ is $N=1$.
$N$ increases linearly with the number of time steps taken.
We can afford to make 
the timesteps sufficiently small to be able
to follow the ionization and energy equations accurately
in this simple approach.

To self-consistently compute the IMF,
we follow the metal enrichment history of the intergalactic medium.
Most importantly, there is a critical transition 
at some point in redshift
in the metallicity of the IGM, when a certain amount of Pop III
stars have formed.
Subsequent to that transition, sufficient metal cooling would cool
collapsing gas inside halos to temperatures lower than achievable
by \h2 cooling, resulting in the ``normal" Pop II stars 
with a ``normal" Salpeter-like IMF.
According to Oh \etal (2001) this transition occurs, when
a fraction, $3\times 10^{-5}-1.2\times 10^{-4}$,
of the total gas is formed into VMS with masses 
in the range $140-260\msun$.
Without fine tuning, it is likely that
VMS with masses outside this range should exist.
Bearing that in mind 
we assume conservatively for our calculations a threshold 
fraction of $1\times 10^{-4}$. 
We note that between the formation of stars and metal enrichment
of the IGM there should be a delay in time.
To accommodate this time lag we assume
$0.2t_H$ (where $t_H$ is the Hubble time at that time)
after a fraction  $1\times 10^{-4}$ of total gas has formed into VMS,
the transition from Pop III VMS to normal
stars with Salpeter-like IMF takes place.

The emission spectrum of metal-free VMS 
with $M\ge 100\msun$ is relatively simple and
has, to a very good approximation,
a blackbody spectrum with 
an effective temperature of $\sim 10^{5.2}~$K,
radiating at the Eddington luminosity
$L_{\rm Edd}=1.3\times 10^{38}(M/\msun)$erg~s$^{-1}$
for about $3\times 10^{6}~$yr,
which translates to 
a hydrogen ionizing photon production rate
of $1.6\times 10^{48}~$photons~s$^{-1}$~$\msun^{-1}$
for VMS 
(El Eid, Fricke, \& Ober 1983;
Bond, Arnett, \& Carr 1984;
Bromm, Kudritzki, \& Loeb 2001;
Schaerer 2002).
The hydrogen ionizing photon rate 
for normal Pop II stars with Salpeter-like
IMF with a low metallicity is
assumed to be $8.9\times 10^{46}~$photons~s$^{-1}$~$\msun^{-1}$,
corresponding to a UV emission efficiency of $1\times 10^{-4}$ 
(in units of the rest mass energy of the total amount of stars formed),
adopted from Bruzual-Charlot 
stellar synthesis library (Bruzual 2000) for metal poor stars.
With the high effective temperature
VMS are efficient emitters of ionizing photons 
with approximately ten times more hydrogen ionizing photons
per unit stellar mass than stars with the Salpeter IMF.

We use a {\it very conservative}
star formation efficiency for minihalos with \h2 cooling
$c^*_{\rm H2}=0.002$ (Abel \etal 1997, 2000; Bromm \etal 2001,2002),
the fraction of gas formed
into stars out of an amount of gas virialized in minihalos.
Star formation efficiency in halos with efficient
atomic cooling is assumed to be $c^*_{\rm HI}= 0.1$.

\subsection{Normalizing Ionizing Photon Escape Fraction}

The most important factors determining
the ionization process are
$c^*$, $f_{\rm es}$, $C$.
The first two can be combined (see Equation 15).
Thus, effectively,
there are two primary factors, 
$c^* f_{\rm es}$ and $C$.
For the sake of explanation and taking
a very conservative stance
we use a ratio $c^*_{\rm HI}/c^*_{\rm H2}=50$,
as indicated in the previous subsection,
to maximize the radiation emission for the second
reionization when Pop II stars are responsible
and minimize the emission from Pop III 
stars that are responsible for the first reionization.
Consequently, we are left with 
the freedom to adjust $f_{\rm es}$ and $C$.

\begin{figure}
\plotone{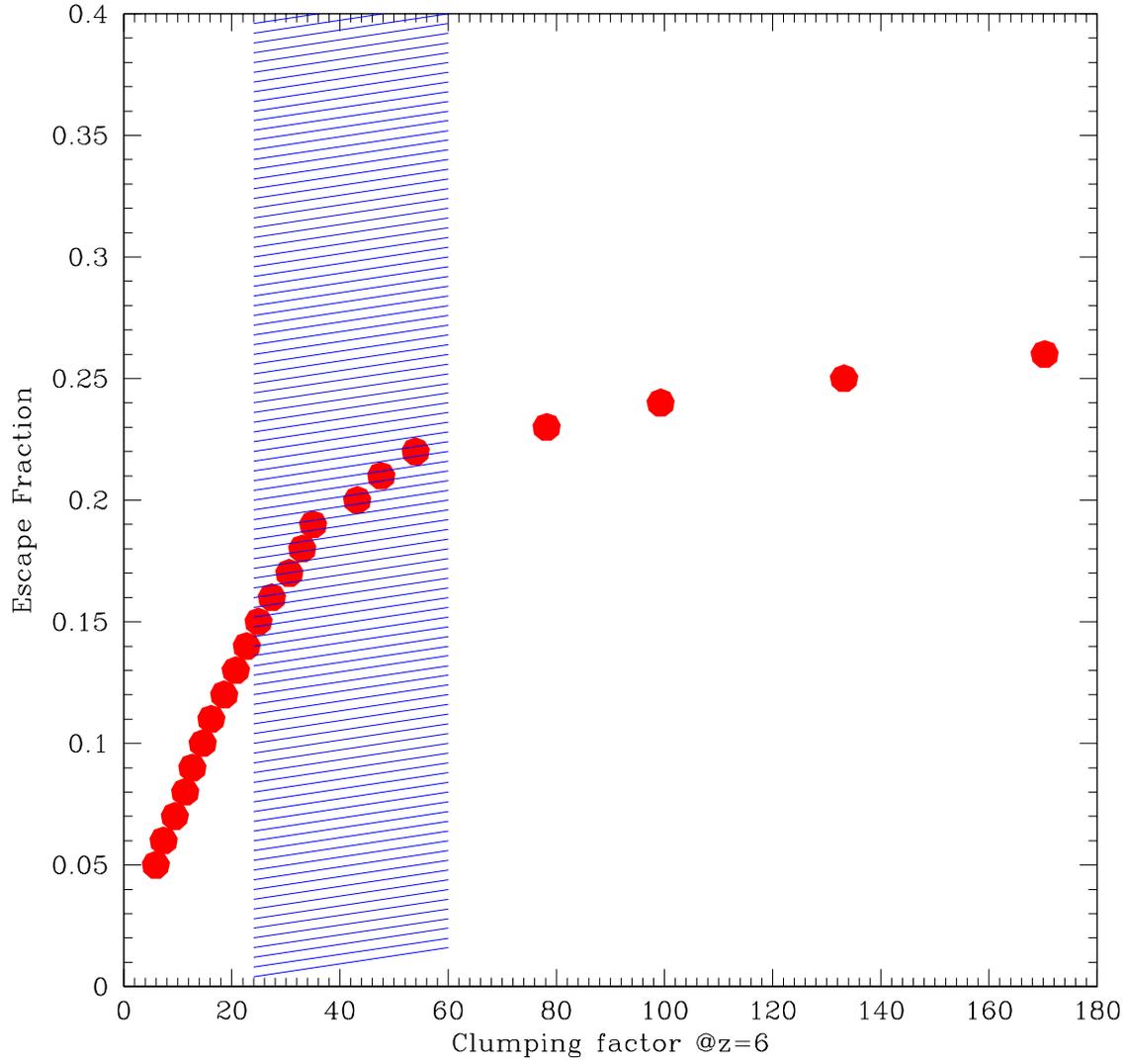}
\caption{
shows the required ionizing photon escape fraction
from galaxies
as a function of the clumping factor of the IGM at $z=6$.
The cross-shaded area indicates the range 
of the clumping factor constrained by \lya
forest observations ($2\sigma$; Croft \etal 2002).
}
\label{time}
\end{figure}

Both theoretically and observationally, we have essentially
no direct knowledge about $f_{\rm es}$ at high redshift in question.
One may argue that smaller galaxies
at high redshift would result in higher escape fraction.
On the other hand, one may also argue that
higher density at high redshift may 
yield lower escape fraction (e.g., Ricotti \& Shull 2000).
It thus seems most productive to seek 
empirical constraints on $f_{\rm es}$.
Our approach is to 
require that the universe is reionized at $z=6$,
as observations suggest
(e.g., Fan \etal 2001; Becker \etal 2001, Barkana 2001; 
Cen \& McDonald 2002; Litz \etal 2002).
With such a normalization point
and the fact that the clumping factor
$C$ is well constrained by \lya forest
observations, we 
are able to tightly constrain the range of $f_{\rm es}$.
Since the same range of waves in the density fluctuation
power spectrum traced by the \lya forest 
are mostly responsible for the clumping of the IGM at $z=6$.
This direct observational constraint is very powerful.

Figure 7 shows the 
required ionizing photon escape fraction
from galaxies
as a function of the clumping factor of the IGM at $z=6$.
The cross-shaded area indicates the range 
of the clumping factor constrained by \lya
forest observations (Croft \etal 2002),
which is obtained as follows.
Based on the latest high signal-to-noise
\lya forest observations,
Croft \etal (2002) give a constraint on 
$\sigma_8=0.78 \pm 0.22$ ($2\sigma$) for a flat cold dark matter
model with $\Omega_M=0.25$ and $\Lambda=1-\Omega_M$.
Using slightly different values of $\Omega_M$ does not significantly
change the range of $\sigma_8$.
We use simulations to empirically obtain the
clumping factor of HII regions at $z=6$.
The high resolution simulations of Gnedin \& Ostriker (1997) 
are the best simulations for this purpose.
They give a clumping factor for HII regions $C_{\rm HII}=35$ at $z=6$
for a low $\sigma_8$ $\Lambda$CDM model with
$\sigma_8=0.67$, $\Omega_M=0.35$, $h=0.70$ (see Figure 2 of 
Gnedin \& Ostriker 1997).
Since the $\Lambda$CDM model at the high redshift
behaves like an Einstein-de Sitter universe
and as a result we can scale their results approximately as follows.
We can obtain the clumping factor 
at $z=6$ for a $\sigma_8=0.56$ model by
using the clumping factor at $z=7.4$ 
of the Gnedin \& Ostriker (1997) simulation
and similarly the clumping factor 
at $z=6$ for a $\sigma_8=1.0$ model by
using the clumping factor at $z=4.0$ of their simulation.
Constraining to have an ionized universe at $z=6$
we obtain $C_{\rm HII}=24-60$ at $z=6$
for the $\Lambda$CDM model
with $\sigma_8=0.56-1.0$ ($2\sigma$),
shown as the cross-shaded region in Figure 7.
The solid dots in Figure 7 are obtained by
running simulations of the reionization process 
using the method described \S 4.1
by requiring that reionization complete at $z=6$
and the clumping factor at $z=6$ matches the value on the x-axis.

Figure 8 shows the epoch of the first reionization 
as a function of the normalized escape fraction.
The cross-shaded region is constrained
by the normalization that the universe is reionized
at $z=6$ (see Figure 7).
We see that, the required constraint by the \lya forest observations
limits the first reionization epoch to $z=15-16$.

\begin{figure}
\plotone{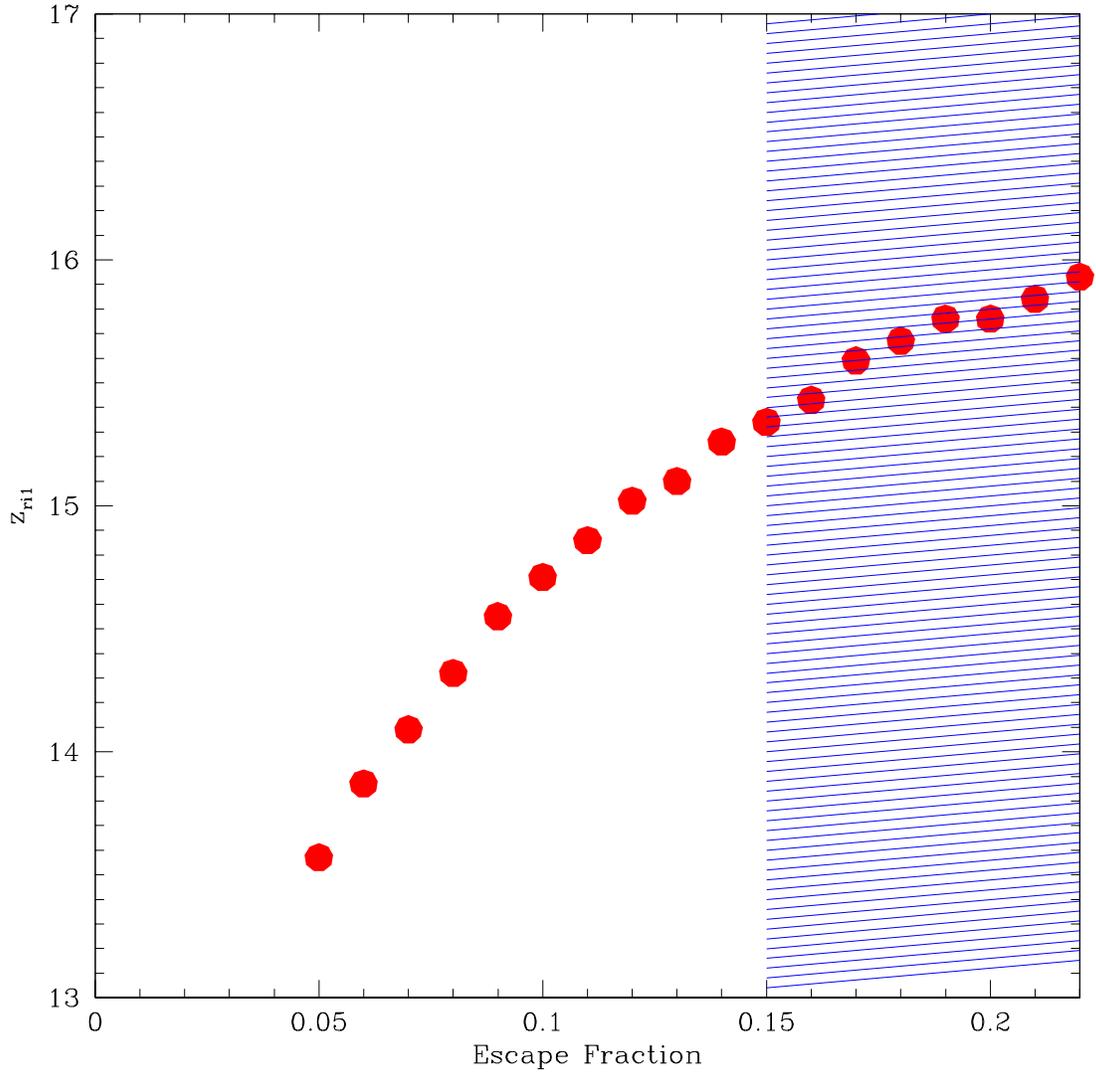}
\caption{
shows the epoch of the first reionization 
as a function of the normalized ionizing photon escape fraction from galaxies.
The cross-shaded region is constrained
by the normalization that the universe is reionized
at $z=6$ (see Figure 7).
}
\label{time}
\end{figure}

We note that the required ionizing photon escape fraction
of $0.15-0.23$ from galaxies at $z\sim 6$
is significantly higher than those
of local starburst galaxies.
It should stressed that the clumping factor predicted 
by Gnedin-Ostriker simulations, though high resolution,
is nevertheless still underestimated.
Thus, the actual required 
ionizing photon escape fraction from galaxies
may be somewhat higher than indicated.
For local starburst galaxies
Hurwitz \etal (1997) give $f_{\rm es}\le 0.032,0.052, 0.11$ ($2\sigma$)
for Mrk 496, Mrk 1267,
and IRAS 08339+6517 ($\le 0.57$ in the case of Mrk 66).
Deharveng \etal (2001) give an escape fraction of $f_{\rm es}<0.062$
for Mrk 54.
Heckman \etal (2001) give $f_{\rm es} \le 0.06$ 
from local starburst galaxies.
Theoretical models (Dove \& Shull 1994;
Dove \etal 2000) give an estimate
of $f_{\rm es}=0.02-0.10$.
Thus, we conclude that, in the context of $\Lambda$CDM model,
galaxies at $z\sim 6$ 
appear to demand a higher ionizing photon escape fraction
than local starburst galaxies.
Upon extrapolation it seems plausible that 
a higher ionizing photon escape fraction
may be expected for galaxies at still higher redshift,
those responsible for the first cosmological reionization.
For simplicity and being conservative
we have assumed that the 
ionizing photon escape fraction 
from galaxies at $z\ge 6$
is equal to that as determined in Figure 7.

\subsection{Detailed Evolution of Intergalactic Medium}

Let us now examine the evolution of IGM
in the context of standard cold dark matter model.
We use the following model parameters:
$c^*_{\rm III}=0.002$,
$c^*_{\rm II}=0.1$,
$<C>=25$ at $z=6$, 
$f_{\rm es}=0.15$.
This model lies at the lower (left) bound
of the cross-shaded region in Figure 7.
Any model to the right of this model would require
higher ionizing photon fraction and thus would
yield a higher first reionization redshift.
A spatially flat cold dark matter cosmological model 
with $\Omega_M=0.25$, $\Omega_b=0.04$, $\Lambda=0.75$,
$H_0=72$km/s/Mpc and $\sigma_8=0.8$ is used.
The results do not sensitively depend on small variations
on the cosmological model parameters.
We remind the reader that in our formalism described
in \S 4.1 we have the freedom to adjust 
$C_{\rm halo}$ to match the required $<C>$ at $z=6$;
we find that 
$C_{\rm halo}=702$ with the adopted cosmological model
provides a match to the required $<C>=25$ at $z=6$.

\begin{figure}
\plotone{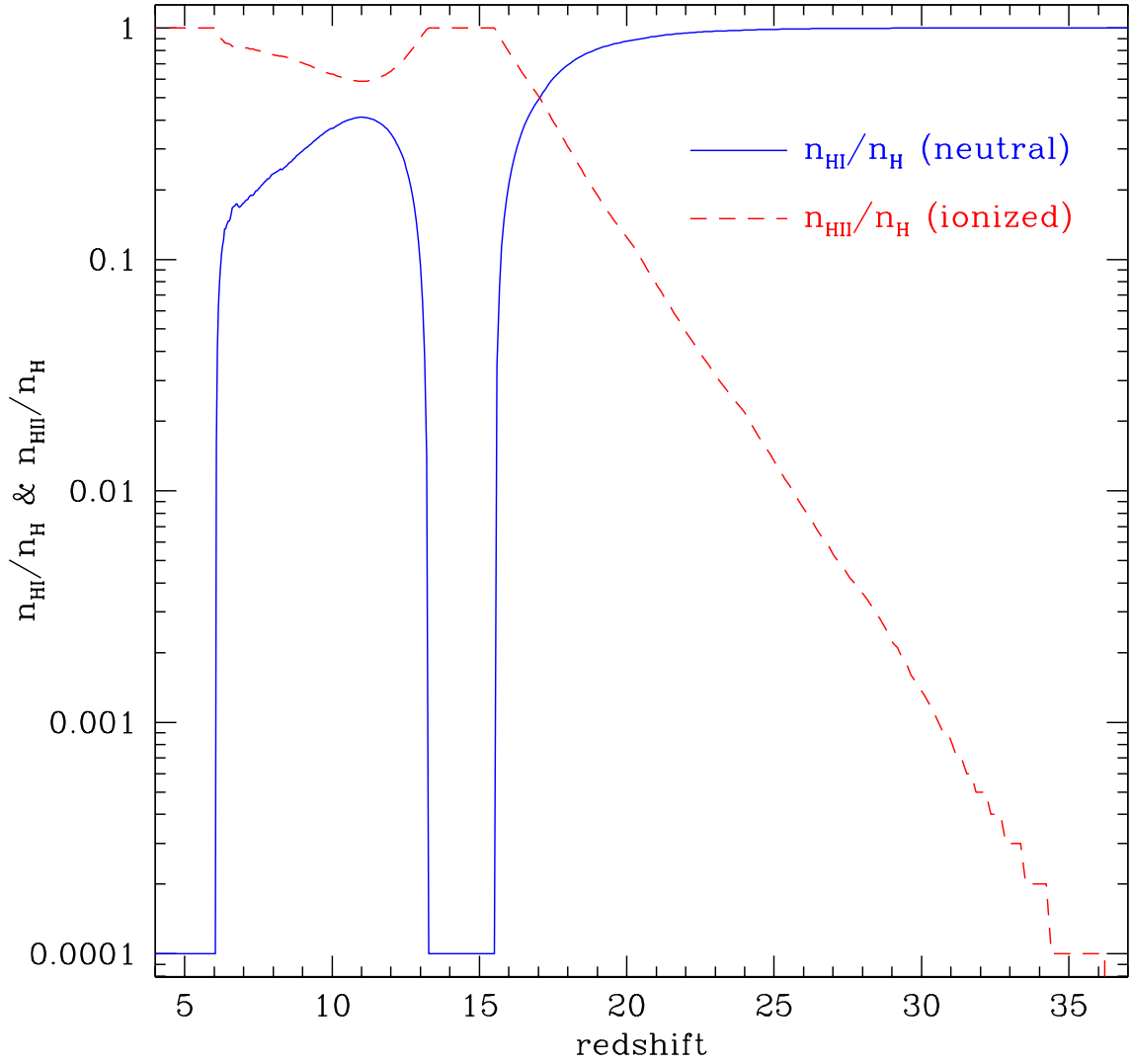}
\caption{
shows the global mean of 
the hydrogen neutral (solid) and complimentary ionized (dashed)
fraction as a function of redshift.
}
\label{y}
\end{figure}

\begin{figure}
\plotone{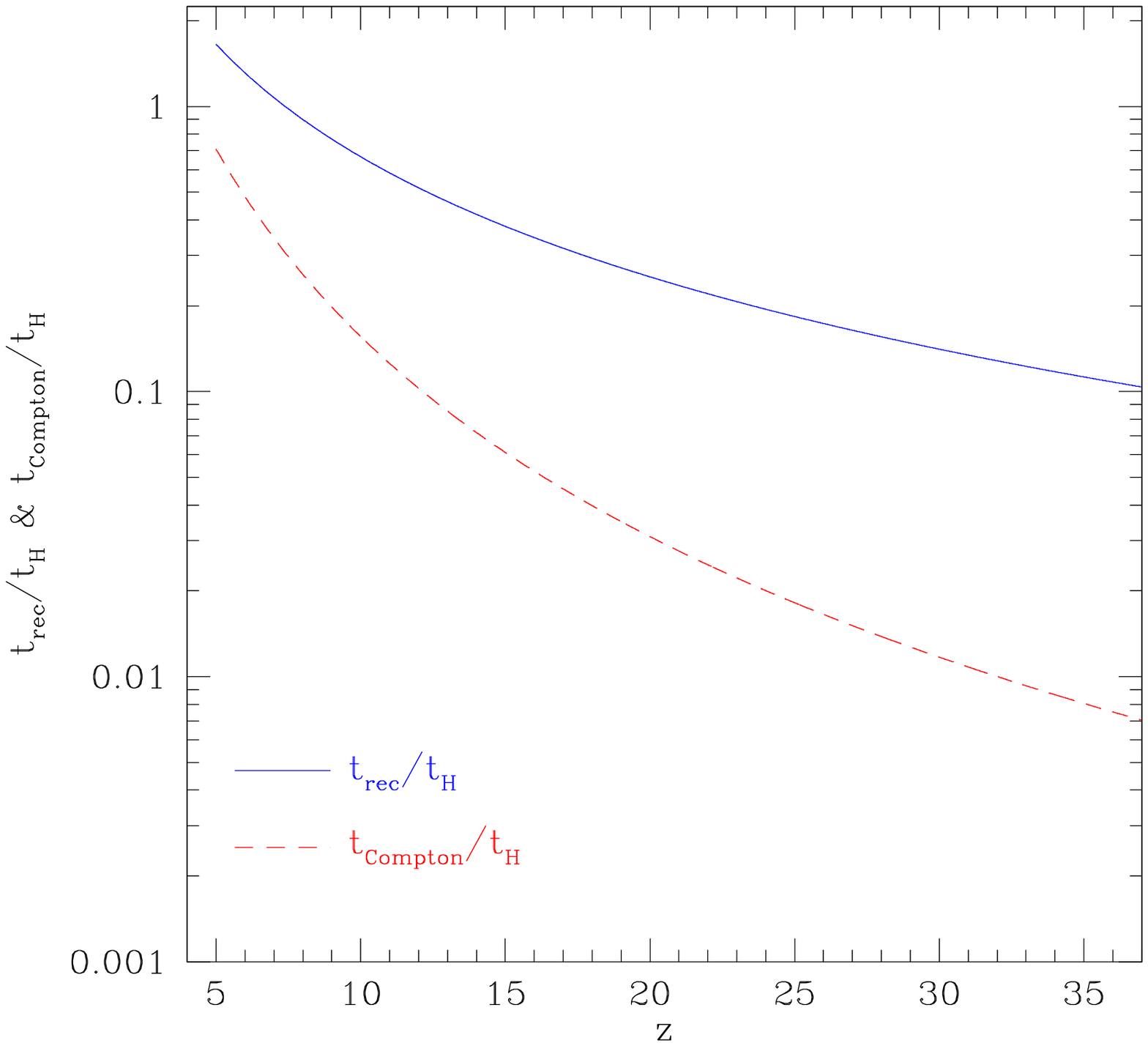}
\caption{
shows the ratio of the recombination time to the Hubble time (solid curve)
and ratio of the Compton cooling time to the Hubble time (dashed curve)
a function of redshift.
}
\label{time}
\end{figure}

Figure 9 shows the global hydrogen neutral fraction and
the complimentary ionized fraction
as a function of redshift.
The first reionization at $z=15.5$ 
as well as the sustained ionized state until 
$z=13.2$ is made possible by Pop III VMS.
The redshift $z=13.2$ marks
the transition from Pop III stars to Pop II stars,
causing the emission of hydrogen ionizing photons to plunge.
The suddenly reduced hydrogen ionizing photon emission
is no longer able to counter the
rapid hydrogen recombination process,
resulting in the second cosmological recombination
at $z=13.2$.
Since a very small amount of neutral fraction suffices
to blank out all \lya emission, the universe
essentially becomes opaque to \lya photons from 
$z=13.2$ to $z=6$.

However, hydrogen is significantly 
ionized with $n_{\rm HII}/n_{\rm H}\ge 0.6$ throughout the long 
second reionization process from $z=13.2$ to $z=6$.
To facilitate a better understanding,
it is useful to show some important time scales involved.
We show in Figure 10 the ratio
of hydrogen recombination time over the Hubble time 
and the ratio of Compton cooling time over the Hubble time.
It is noted that at the redshift and density of interest 
Compton cooling dominates over other cooling terms,
although adiabatic cooling starts to become important
approaching the end of the second reionization period.
We immediately see that the hydrogen recombination time is significantly
longer than the Compton cooling time,
both of which are shorter than the Hubble time at $z\ge 8$.
Therefore, the IGM at early times heated up by the 
photoionization would subsequently cool down more rapidly than recombining,
resulting in overcooled but significantly ionized IGM,
in the absence of subsequent photoheating.

\begin{figure}
\plotone{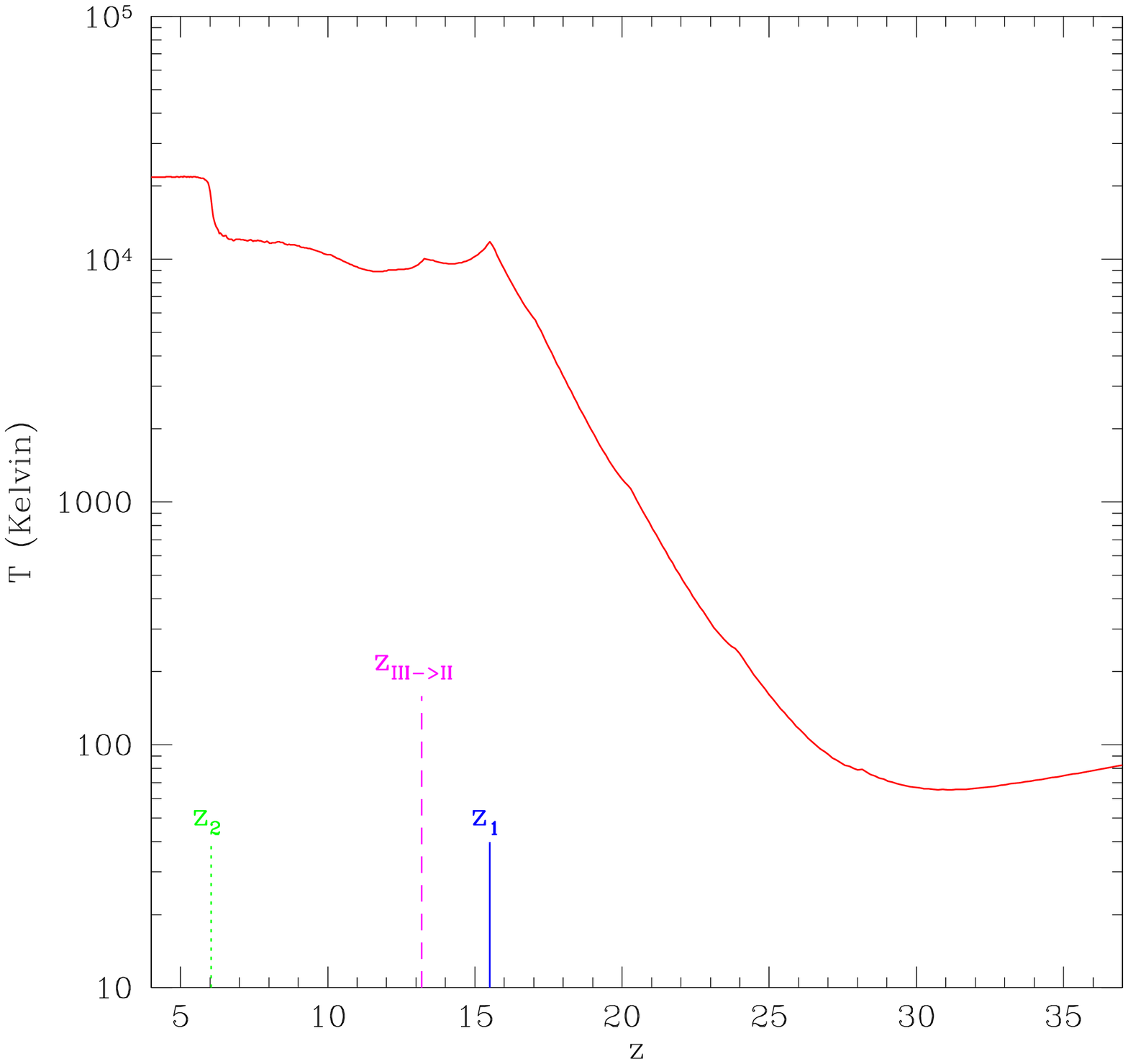}
\caption{
shows the evolution of the mean IGM temperature 
as a function of redshift
during the second cosmological reionization process.
The solid vertical tick indicates the first reionization epoch.
The dashed vertical tick indicates the transition epoch 
from Pop III stars to Pop II stars.
The dotted vertical tick indicates the second reionization epoch.
}
\label{tem}
\end{figure}

\begin{figure}
\plotone{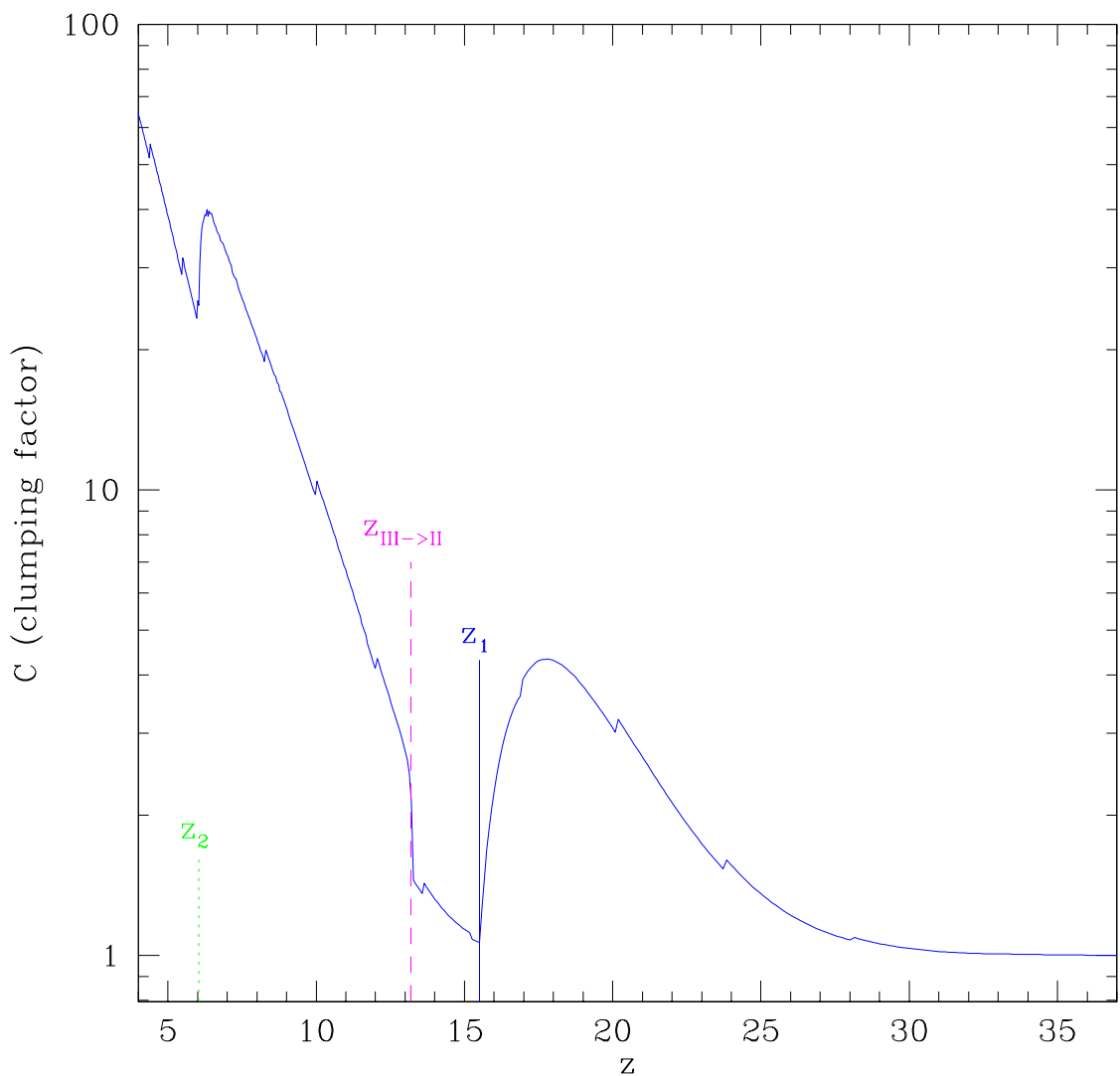}
\caption{
shows the evolution of the mean clumping factor 
as a function of redshift
during the second cosmological reionization process,
consistent with direct numerical simulation of
Gnedin \& Ostriker (1997).
The solid vertical tick indicates the first reionization epoch.
The dashed vertical tick indicates the transition epoch 
from Pop III stars to Pop II stars.
The dotted vertical tick indicates the second reionization epoch.
}
\label{clumping}
\end{figure}

\begin{figure}
\plotone{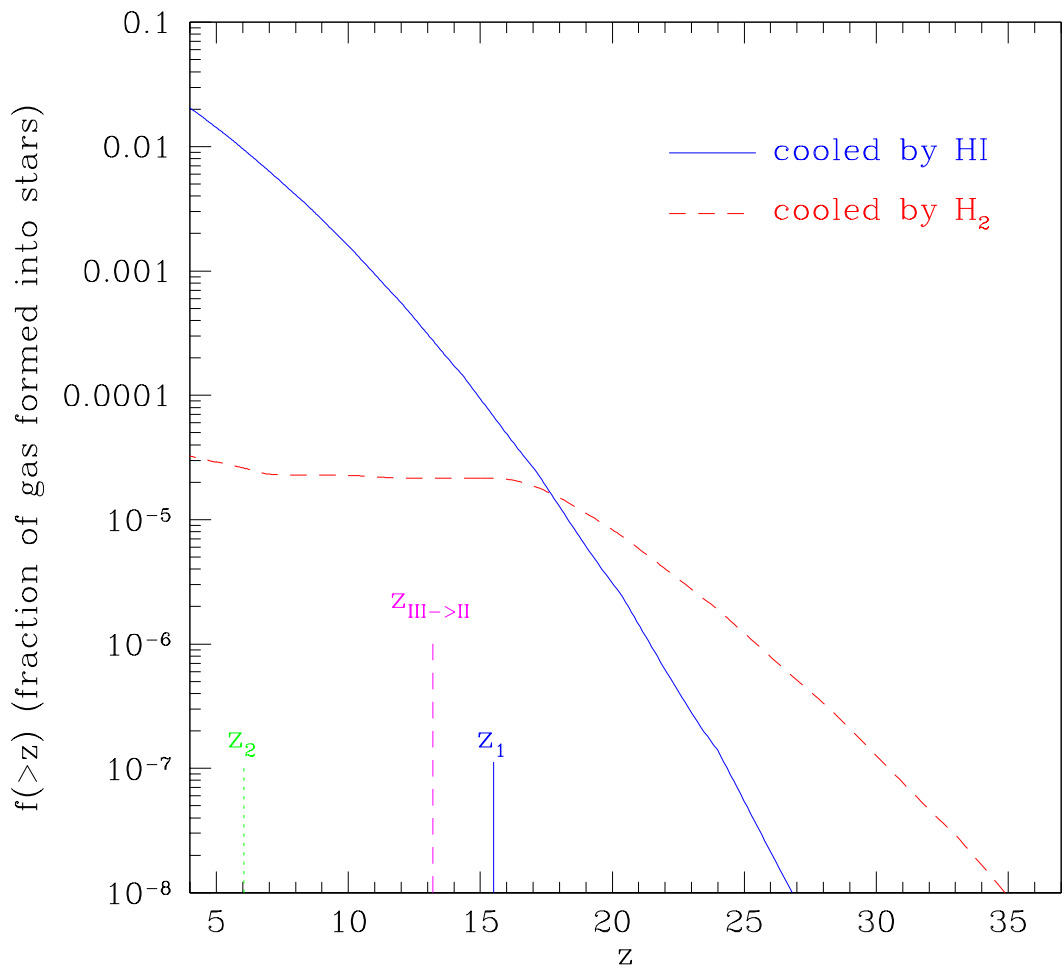}
\caption{
The left y-axis shows the evolution of the fraction of gas formed
into stars as a function of redshift.
The right y-axis translates 
the fraction of gas formed
into stars into the number of ionizing photons
per baryon as a function of redshift.
The solid vertical tick indicates the first reionization epoch.
The dashed vertical tick indicates the transition epoch 
from Pop III stars to Pop II stars.
The dotted vertical tick indicates the second reionization epoch.
}
\label{gal}
\end{figure}

\begin{figure}
\plotone{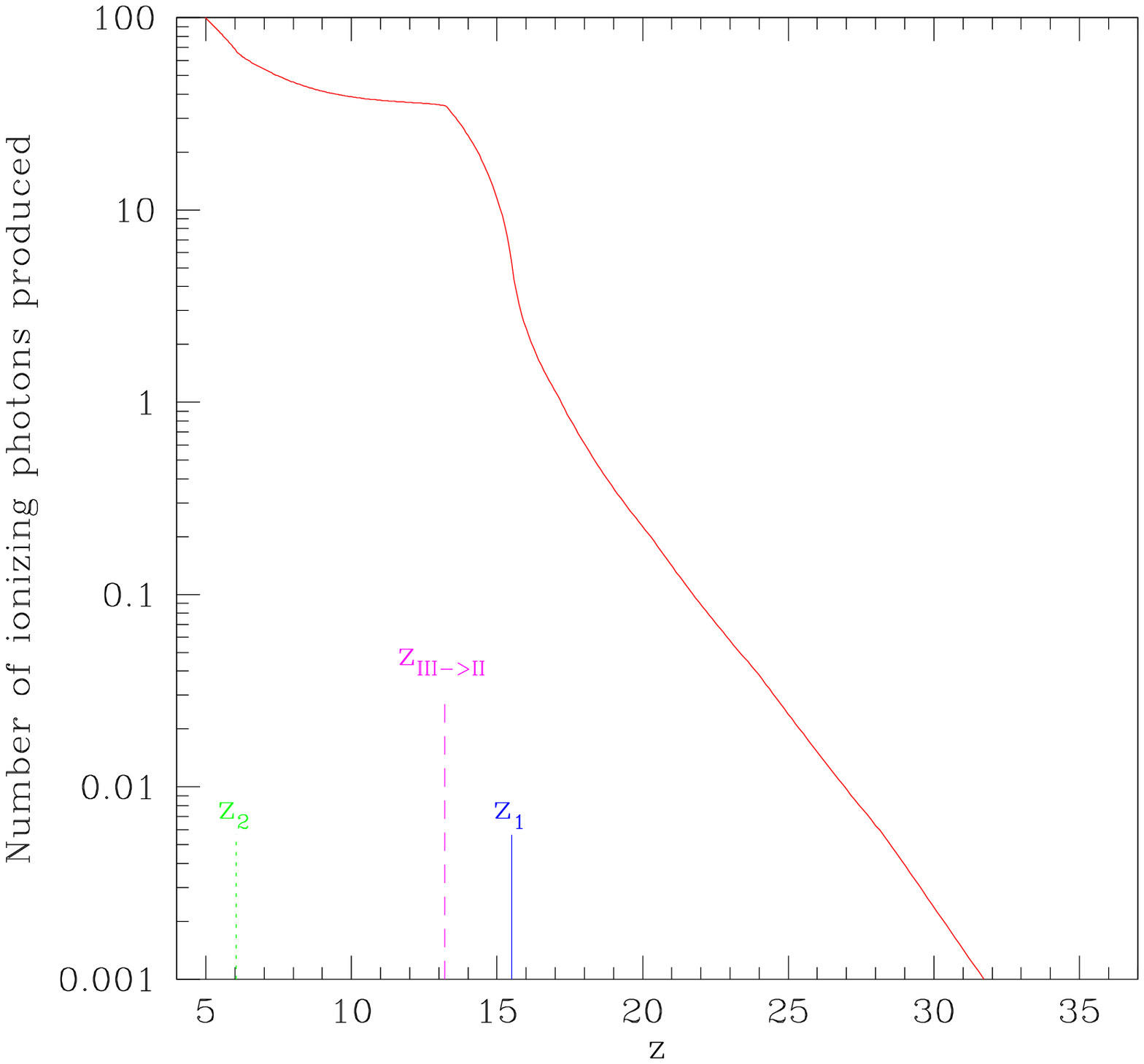}
\caption{
shows the cumulative number of ionizing photons per hydrogen atom
produced as a function of redshift.
The solid vertical tick indicates the first reionization epoch.
The dashed vertical tick indicates the transition epoch 
from Pop III stars to Pop II stars.
The dotted vertical tick indicates the second reionization epoch.
}
\label{gal}
\end{figure}

More results are shown in Figures (11,12,13,14)
for the evolution of mean temperature,
clumping factor, stellar fraction and number of 
ionizing photons, respectively, as a function of redshift.
In Figure (11) we see a sustained ascent 
in mean temperature of the IGM from $z\sim 30$ up to the redshift
of the first reionization $z=15.5$.
Subsequently, the mean temperature of the IGM
is roughly maintained at $10^4$K, due to
the works of two counter-balancing effects:
substantial cooling reduces
the Jeans mass of the IGM and would
increase the star formation rate, which would then
provide increased heating by photoionization.
Thus, the mean temperature of the IGM is
self-regulated due to the competition
between the Compton cooling and photoheating.
For the cold dark matter cosmological model
it happens that
the star formation activities are at a level such that
the two balancing terms are comparable in magnitude,
resulting in a fairly mild evolution 
of the IGM mean temperature seen in Figure 11.

From Figure 12 it is evident that
in the period preceding the first reionization
from $z=18$ to $z=15.5$
a large fraction of gas 
is unable to accrete onto minihalos due to heated temperature
of the IGM and results in
a large decrease in the clumping factor of the IGM,
followed by a mild increase in the clumping factor in the brief 
ionized state from $z=15.5$ to $z=13.2$.
From $z=13.2$ to $z=6$ the 
clumping factor of the IGM 
increases steadily, parallel to the increase in the nonlinear
mass scale with time,
which is then followed a very brief, moderate drop
immediately after $z=6$ when
there is an increase in the temperature of the IGM
from $T\sim 10^4$K to $T\sim 2\times 10^4$K (see Figure 11)
due to the completion of the second reionization.

Some interesting features may be found in Figures (13,14).
We see in Figure 13 that, although stars formation
in minihalos (dashed curve)
dominate over that in large halos (solid curve)
at $z\ge 17$, 
the latter gradually takes over and become dominant
at $z\le 17$.
Therefore, while Pop III stars in minihalos 
may be a very important source of heavy elements
to enrich the IGM,
the first cosmological reionization is largely 
due to Pop III stars in large halos.
As a result, at least for this particular case,
the first reionization would have occurred
even without 
the validation presented in \S 2,3 that \h2 cooling is sufficient
in minihalos during the first reionization process.
About $6\times 10^{-5}$ of the total gas has formed into Pop III VMS
by the time the universe was first reionized.
We note that we have adopted a very conservative
ratio of $c^*_{\rm HI}/c^*_{\rm H2}=0.1/0.002=50$ for this example.
If, for example, the ratio were $10$,
then Pop III stars in minihalos would largely
be responsible for first reionization.
By the time the universe was reionized for the second time.
about $1\%$ of the total gas has formed into stars.
From Figure 14 we see that 
at the epoch of the first reionization
about 4 ionizing photons per baryon have been produced
by Pop III stars.
About $10^{-4}$ fraction of gas formed into Pop III stars
at a slightly later epoch and the delayed transition
from Pop III stars to Pop II stars occur at $z=13.2$,
clearly seen in Figures (9,11,12).
More than 60 ionizing photons per baryon have been produced 
by the time than the universe is reionized for the second
time at $z=6$.

In summary, the cosmological reionization process has several
interesting characteristics.
First, the universe was generically reionized twice, one
at $z=15-16$ by Pop III VMS and the other at $z\sim 6$ by 
normal Pop II stars.
Most of the Pop III VMS responsible for the
first cosmological reionization would come from 
large halos if star formation efficiency in minihalos
is less than a few percent of that of large halos;
in this case, our argument that \h2 cooling is efficient in minihalos
during the first cosmological reionization, while still
true, may not be necessary.
Otherwise, 
most of the Pop III VMS responsible for the
first cosmological reionization would come from 
minihalos, where efficient \h2 cooling is required.
Second, the IGM is maintained at a rather ``warm" temperature
$T\sim 10^4~$K from the first reionization through
the second reionization; i.e., the universe is significantly
heated up even in this ``dark age" preceding the second, complete
reionization of the universe at $z\sim 6$.
Third, the IGM is kept at a significantly
ionized state throughout the second reionization period.
Finally, the overwhelming fraction of star formation
activity responsible for producing the ionizing photons
in the period from the first reionization through the second reionization
are in large halos.

\section{Discussion}

The possible indication of a more top-heavy initial stellar
mass function (IMF) in early galaxies or earlier stages
of galaxies than present-day IMF (Salpeter 1955)
was suggested five decades ago (Schwarzschild \& Spitzer 1953).
Cosmological consequences of 
the first generation, massive stars at high redshift
have been discussed in a variety of contexts
(e.g., Layzer \& Hively 1973;
Carr 1977;
Rees 1978;
Rowan-Robinson, Negroponte, \& Silk 1979;
Puget \& Heyvaerts 1980;
Tarbet \& Rowan-Robinson 1982;
Carr, Bond, \& Arnett 1984; Haiman \& Loeb 1997;
Barkana \& Loeb 2001).

The reionization picture presented here
would have a wide range of profound implications 
for many aspects of structure formation.
There are many questions that need to be addressed.
It is beyond the scope of this paper to 
explore any of these issues in significant details
and we will study these and other relevant issues in 
subsequent investigations.
But we will give some simple estimates or analyses for some selected issues.

\subsection{Initial Metal Enrichment of the Intergalactic Medium}

Direct observational evidence for massive Pop III stars 
may be just emerging recently.
While Luck \& Bond (1985) and others 
have previously indicated the need
for VMS ($M>100\msun$)
to explain the overabundant $\alpha$ elements, 
Qian, Wasserburg and collaborators (Wasserburg \& Qian 2000;
Qian \& Wasserburg 2000,2002;
Oh \etal 2001)
have recently stressed the unique signature of Pop III VMS
and suggested that Pop III stars
could promptly produce the observed abundance patterns
of metal poor stars 
(McWilliam \etal 1995;
Ryan, Norris, \& Beers 1996;
Rossi, Beers, \& Sneden 1999;
Norris, Ryan, \& Beers 1997,2001;
Norris \etal 2002;
Hill \etal 2002;
Depagne \etal 2002). 

Given the shallow potential wells of Pop III galaxies,
supernova explosions would likely
blow away the ejectas along with 
a large fraction of the diffuse interstellar gas 
(Mac Low \& Ferrara 1999; Mori, Ferrara \& Madau 2002).
From Figures (9,11-14) we see that 
the transition from Pop III stars to Pop II stars
occurs a later time than the first reionization;
i.e., Pop III stars ionized the universe for the first time.
Oh \etal (2001) have suggested this possibility
based on the observed transition
of abundance pattern at [Fe/H]=-3 in metal poor stars
summarized by Qian \& Wasserburg 
(Wasserburg \& Qian 2000; Qian \& Wasserburg 2000,2002), 
who found that 
at lower value VMS dominates the enrichment,
whereas at higher value 
a sharp rise in the abundances of the heavy r-process elements
such as Ba and Eu in galactic halo stars with [Fe/H]$\ge -3$
signifies the occurrence of Type II supernovae
of normal stars with masses of $10-60\msun$.

One may expect to see the Pop III star abundance patterns 
in low density regions of the universe,
such as \lya clouds or voids,
where subsequent additional enrichment may be minute.
Recent work on \lya clouds suggests that 
metal enrichment by Pop III stars 
be consistent with \lya cloud observations
and the necessary enrichment occur prior to $z\sim 4.6$
(the highest epoch for the available \lya cloud data analyzed)
(Qian, Sargent, \& Wasserburg 2002).

One obvious advantage for
Pop III supernovae to enrich
the IGM with the first metals at $z\gg 4$
is that it is much easier to relatively uniformly
disperse the metals across the IGM
for two reasons.
First, the disturbances to the density and velocity fields
are smaller at high redshift,
because each Pop III galaxy needs to fill only 
a very small IGM volume.
Second, any significant large-scale motions
would decay away rapidly, in the absence 
of dynamical support.
As a result, the excellent agreement 
found between simulations and
observations with respect to the \lya forest
at $z=2-4$ 
(Cen \etal 1994; Zhang \etal 1995;
Hernquist \etal 1996; Miralda-Escud\'e \etal 1996; Bond \& Wadsley 1997;
Theuns \etal 1998)
would not be altered,
although recent simulations also indicate
that low redshift galactic winds would
not spoil the main properties of
\lya forest produced by previous simulations
(Theuns \etal 2002).

\subsection{Intergalactic Magnetic Field}

Another possible consequence of supernova explosion
and expulsion of the gas into the IGM 
is the magnetic feedback to the IGM.
Rees (1994) has first pointed out the importance
of resident supernova remnants 
in galaxies being a substantial large-scale seed field for galactic dynamos.
Let us give a simple estimate here for the contribution 
of the Pop III supernova remnants to the intergalactic 
magnetic field.
Folloing Rees (1994)
we should use local observations as a guide.
The Crab Nebula (a plerion) has a magnetic field of strength
$B \sim 100\muG$ currently occupying
a sphere of radius $r \sim 1~$pc,
totaling a flux of $\psi_{\rm Crab} \sim 3\times 10^{33}~$G~cm$^2$.
Assuming flux conservation and Pop III magnetic bubble filling 
factor of unity give
$B_{\rm IGM}=({f_{\rm III} \rho_b (1+z_{\rm ej})^3\over M_{\rm III}})^{2/3} 
({M_{\rm III}\over M_{\rm Crab}}) \psi_{\rm Crab}$,
where $M_{\rm III}\sim 100\msun$ is the mass of a typical
Pop III star, $M_{\rm Crab}\sim 5\msun$ 
is the zero-age main sequence
progenitor star for Crab;
$f_{\rm III}\sim 2.5\times 10^{-4}$ is the fraction
of baryons formed into Pop III stars (see Figure 14) by
redshift $z=13.2$ when the transition from Pop III to Pop II occurs;
$\rho_b$ is the mean baryonic density at zero redshift;
$z_{\rm ej}$ is the redshift of ejection of the magnetic field
into the IGM;
we have assumed that the magnetic flux is approximately 
proportional to the stellar mass.
Inserting all the numbers gives
$B_{\rm IGM}(z=13.2) \sim 1\times 10^{-9}~$G
at $z=13.2$ and subsequently
$B_{\rm IGM}(z) \sim 1\times 10^{-9}({1+z\over 14.2})^2~$G.
This magnetic field,
having a significantly larger amplitude
than that produced by gravitational shocks 
in the collapse of large-scale structure (Kulsrud \etal 1997),
could serve as a seed field for subsequent galaxy formation.
The mean separation between Pop III galaxies 
is of order $100$ comoving kpc.
Thus, if Pop III supernovae are responsible
for enriching the IGM relatively uniformly
to a metallicity of about a thousandth of the solar value,
it is also likely that the magnetic field lines originating
from supernovae remnants
would be stretched to fill up the intergalactic space,
resulting in an initial magnetic field 
possibly coherent on scales as large as $\sim 100$kpc.

In addition, miniquasars powered by Pop III
black holes may produce mini radio jets, 
as would have been implied by
the observational fact
that radio jets are observed in accretion disks
on a wide range of scales in a wide variety of astrophysical systems.
The magnetic field from miniquasars may be 
as important as or more important than
that from the Pop III stars;
but a reliable estimate is difficult.

\subsection{Pop III Black Holes}

Without fine tuning the IMF of Pop III stars,
it seems likely that $\sim 200\msun$ black
hole from Pop III stars more massive
than $260\msun$ and smaller black holes ($M_{\rm BH}\sim 10-50\msun$)
from Pop III stars less massive than $140\msun$ would form.
The possible bimodality of the distribution
of black hole mass is interesting
but the consequences are too complicated to 
be easily outlined.
Those Pop III black holes will be building blocks  
for subsequent structures and
the question is: where will they go?
In general, since there is no correlation between
the small-scale structures that form Pop III galaxies
and later structures, 
one would expect to find those black holes
in all environments, including globular clusters,
galactic disks, galaxy halos and intergalactic space.
It is possible that the dynamic formation of 
halos and galaxies in hierarchical structure
formation may be significantly altered in the
presence of these massive black holes,  
especially in the cores of the relevant structures,
such as globular clusters, galaxies, etc.
The number density of Pop III black holes 
of mass $\sim 100\msun$ is of order
$10^3-10^4$ per comoving Mpc$^3$,
comparable to the number density of globular clusters 
and mass density comparable to present mass density
in supermassive black holes in the centers of galaxies
(e.g., Merritt \& Ferrarese 2001).
The interesting agreement between this result and
that of Madau \& Rees (2001) with regard to the 
mass density of Pop III black holes
is traceable to the fact that the $3\sigma$
density peaks used in Madau \& Rees's (2001) model happen
to yield about the same collapsed fraction as ours. 
From the inferred number density of Pop III black holes
it is quite possible that many globular clusters
could be seeded by those black holes.
There is clearly no shortage of 
Pop III black holes to provide seeds for later supermassive
black holes seen in centers of local galaxies (e.g., Tremaine \etal 2002),
AGN and quasars (e.g., Rees 1984,1990).

How could these black holes be observed?
Miralda-Escud\'e \& Gould (2000) pointed out
the possible existence of stellar black hole clusters 
at the Galactic center.
Their reasoning may be equally applied here.
Since Pop III star mass fraction
is of order of $10^{-4}$ of total baryonic mass,
as our calculation indicates,
their should be $\sim 10^3$ black holes 
either having mass $100\msun$ or $10\msun$ 
(noting that the mass fraction in black holes 
will be a factor about 10 lower for less massive 
black holes).
The reader is encouraged to refer
to Miralda-Escud\'e \& Gould (2000) 
and also Sigurdsson \& Rees (1997)
for a detailed discussion of possible manifestations of such
a black cluster, bearing in mind that the black holes
here are more massive and may enhance some of the effects
discussed.

In the hierarchical structure formation process, 
these Pop III black holes are expected to grow
(there are a large number of e-folding times available)
and merge, along with larger cosmic structures.
The merger of black holes may be detectable by 
the gravitational wave experiment Laser
Interferometer Space Antenna (LISA)
(e.g., Menou, Haiman, \& Narayanan 2001; Hughes 2002).

\subsection{Detecting Pop III Galaxies and Pop III Hypernovae}

We will examine the observability
of Pop III galaxies and their associated hypernovae
with respect to SIRTF and JWST (James Webb Space Telescope).

The physical sizes of the galaxies in the redshift range $z=13-15$
is unknown but it seems unlikely that they will be
larger than their lower redshift counterparts.
Steidel \etal (1996) find that 
the great majority of $z>3$ objects have a half-light radius 
$\sim 2$~kpc, which, if placed at $z_S=17$, would
subtend $\sim 0^{''}\hskip -3pt .05$ (for the adopted model), 
more than a factor of $\sim 10$ below the angular resolution of SIRTF.
In other words, these high redshift galaxies are 
point sources to SIRTF.
The flux density (the power per unit antenna
area and per unit frequency interval)
of a point source (without cosmological surface brightness dimming)
at frequency $\nu$ is (Weinberg 1972):
$S={P(\nu [1+z_{\rm S}])\over(1+z_{\rm S})^3 d_{\rm cm}^2}$,
where $z_S$ is the source redshift and $\nu$ is the observed photon
frequency.
$P$ is the intrinsic power, the power emitted per unit solid angle
and per unit frequency interval,
related to the luminosity $L$ of the source 
by $P=L/4\pi\Delta\nu$,
where $\Delta \nu$ is the rest frame bandwidth at which
the source has a luminosity $L$;
$d_{\rm cm}$ is the comoving distance to the source.
Using a rest frame band $0.09-0.27\mu m$ 
(corresponding to the observer's band $1.3-3.8\;\mu m$)
and $z_{\rm S}=13.2$ we obtain
\begin{equation}
S= 1.45 ({L\over 10^9 L_\odot})~\hbox{nJy}. 
\end{equation}
\noindent 
The point source detection sensitivity at $1\sigma$ level
(for relatively long integration time $>500\;$seconds) of 
the Infrared Array Camera (IRAC) on SIRTF 
at $3.5-4.5\;\mu m$ is (SIRTF Project 1997) 
$S_{3.5}\approx \sqrt{(100/t) + 0.6^2}\;\mu$Jy,
where $t$ is the integration time in seconds.
The second term inside the square root 
is the confusion noise limit due to faint unresolved sources,
modeled by Franceschini \etal (1991).

The total luminosity of a Pop III galaxy ranges
from 
$L=1\times 10^7 (c_{\rm H2}^*/0.002)(M_{\rm h}/10^6\msun) \lsun$ for minihalos
to
$L=1\times 10^{10} (c_{\rm HI}^*/0.1)(M_{\rm h}/2\times 10^7\msun) \lsun$
for large halos;
only a small fraction of the total luminosity is in the indicated band.
It is clear that no Pop III galaxies hosted by minihalos
will be directly detectable by SIRFT, even 
without the confusion noise limit.
Perhaps only the high mass end of Pop III galaxies with
mass $\ge 10^{9}\msun$ (which would give $\ge 0.7\mu$Jy)
may be detectable by SIRTF, in the presence of confusion sources.
The sensitivity of the JWST
is close to $1$nJy (Stockman \& Mather 2000).
Therefore, Pop III galaxies with large halos 
will be detectable by JWST, whereas 
Pop III galaxies with minihalos may still be beyond reach of even JWST.
Our calculations indicate a large fraction 
of Pop III galaxies will be detectable by JWST;
a more detailed 
treatment will be reserved for a future paper.

The total number of Pop III hypernovae (HN) is
$N_{\rm HN}=2.5\times 10^{-5} \rho_b ({4\pi/3}) R_{\rm H}^3/M_{\rm III}
=1.3\times 10^{16}$,
where $M_{\rm III}=100\msun$ for Pop III star mass is used,
$R_H=6000h^{-1}~$Mpc is the comoving Hubble radius
and $\rho_b$ is the comoving mean baryonic density.
We can estimate the Pop III
hypernova surface number density at any given time 
to be
\begin{equation}
\Sigma_{\rm HN}=1.3\times 10^{16}\times \left({\Delta t_{\rm HN}(1+z) \over t_{\rm H}}\right)\left({1\over 4\pi\times 1.2\times 10^7 \hbox{arcmin}^2}\right) \times 0.60 \approx 2.3~\hbox{arcmin}^{-2},
\end{equation}
\noindent
where $\Delta t_{\rm HN}$ is the time duration of each hypernova event
in the hypernova restframe and $t_{\rm H}$ is the age of
universe at $z=13.2$;
the first parenthesized term in the above equation
takes account the finite duration of each supernova event
and an intrinsic duration for the hypernovae $\Delta t_{\rm HN}=1~$yr is used;
the last term $0.60$ takes into account that the observable redshift
interval $z\ge 13.2$ is $60\%$ of the total volume of the universe then. 
This HN surface density is comparable to the
surface density of SNe derived by 
Miralda-Escud\'e \& Rees (1997),
based on the observed metallicity in the \lya clouds.

Assuming the optical luminosity of a hypernova to be 
$1\times 10^{10}\; L_\odot$ (about $10$ times that of 
the local normal supernovae),
the flux density of Pop III hypernovae at $z=13.2$
will be at a level $\sim 0.014\mu\hbox{Jy}$. 
Even without background confusion 
({\it if the variable source has a variability time scale shorter than 
the indicated integration time}),
in order to detect such a flux density at $1\sigma$ level,
an integration time of $6~$days on SIRTF IRAC will be required,
which seems impractical.
Even if this long integration is achievable,
the fact that the variability time scale of a hypernovae 
at the considered redshift may be quite long 
indicates that it will be extremely difficult
to directly detect high redshift hypernovae,
{\it unless their flux is significantly amplified}.

\begin{figure}
\plotone{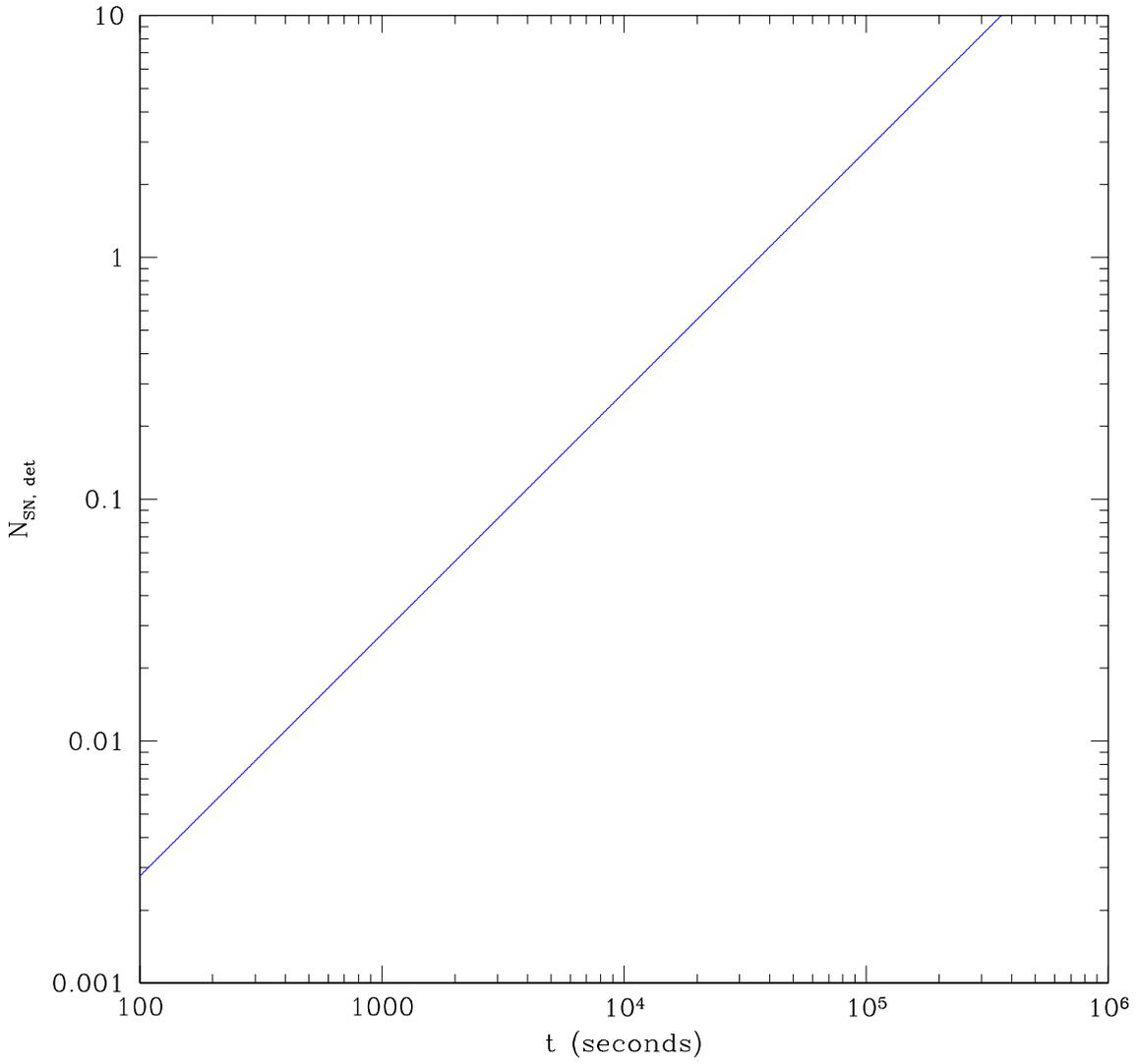}
\caption{
shows the average number of strong gravitational
lensing magnified Pop hypernovae 
that will be observed per IRAC field of view ($25~$arcmin$^2$)
as a function of integration time.
}
\label{Nt}
\end{figure}

In Appendix A we derive the probability 
of strong gravitational lensing by massive clusters of galaxies
and find that a random source behind a massive galaxy
cluster will have a probability 
\begin{equation}
P_{\rm clust}(\mu)={6.7\times 10^{-2}\over \mu^2}\left({\Omega_{\rm FOV}\over 25~\hbox{arcmin}^2}\right)^{-1}
\end{equation}
\noindent 
being magnified by $\ge \mu$,
where $\mu$ is the lensing magnification
and $\Omega_{\rm FOV}$ is the field of view of a telescope.
The number of Pop III hypernovae, magnified by $\ge \mu$, per field of view
centered on a massive cluster is then
\begin{equation}
N_{HN} = P_{\rm clust}(\mu) \Omega_{\rm FOV} \Sigma_{\rm HN}.
\end{equation}
\noindent 
In order for the IRAC camera of SIRTF
to detect at $n\sigma$ statistical level
a point source, which would have
a flux density of $S$ in the absence of gravitational lensing magnification, 
one requires that
\begin{equation}
S \mu = n \sqrt{100/t},
\end{equation}
\noindent 
where point source confusion limit is removed;
$t$ is integration time in seconds.
We plot the number of hypernovae,
$N_{\rm HN, det}$, per field of view detected at $3\sigma$ confidence
level against integration time $t$ in Figure 13,
assuming $L_{\rm HN}=1\times 10^{10}\lsun$ 
giving unlensed flux density of $S=14.5\hbox{nJy}$ 
at the observed wavelength of several $\mu m$.
To put the matter in perspective,
1,000 target fields each with $10$k-second integration time
would detect $\sim 300$ multiply lensed
Pop III hypernovae at $3\sigma$ confidence level.

We note that since only highly magnified Pop III hypernovae
will be observable,
the observed candidates should have distinct features: 
the multiple images will significantly help
the identification process.
If the cluster lens is known, its lens potential
derived independently elsewhere would
further constrain the image configurations.
Once such images are found, periodic 
monitoring up to a few years would eventually
verify the transient nature of the Pop III hypernovae.
Since the total number of detectable hypernovae
is just proportional to the total integration time,
one could choose an optimal strategy
such that the observation is most sensitive
to the expected splittings of the images:
a shorter exposure will require a higher
magnification from a more central region of the lens,
where splittings may decrease due to a smaller velocity
dispersion at the center of the cluster.

At a flux density of a level $\sim 10$nJy,
Pop III hypernovae
will be detectable by JWST (Stockman \& Mather 2000) 
without gravitational lensing magnification.
The possible observable duration of $4-5$yrs for the initial
bright phase of Pop III hypernovae,
corresponding to $80-90~$d intrinsic duration  
(Woosley \& Weaver 1996) will be a signature.

\subsection{Effects on the Cosmic Microwave Background}

\begin{figure}
\plotone{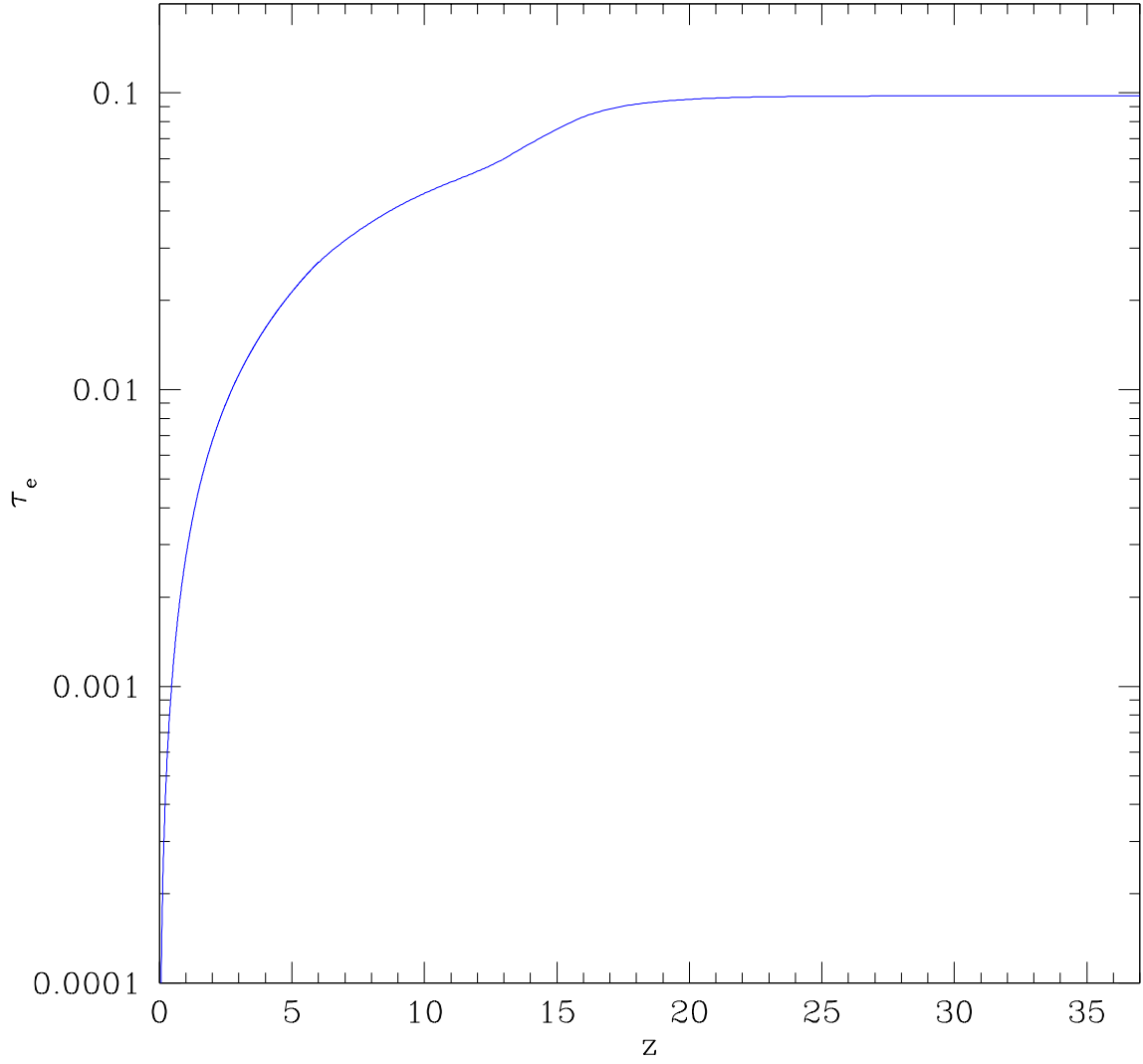}
\caption{
shows the cumulative Thomson scattering 
optical depth as a function of redshift.
}
\label{tau}
\end{figure}

The earlier reionization would significantly increase the electron Thomson 
scattering optical depth.
This is so because the IGM ionization fraction
remains appreciable during the period
between the first reionization ($z=15.5$)
and the second reionization ($z=6.1$).
Figure 16 shows the cumulative Thomson scattering 
optical depth as a function of redshift.
We see that the Thomson scattering optical depth
increases from $\tau_e(z<6.1)=0.027$ to $\tau_e(z<20)=0.097$.
Evidently, the first reionization boosts
the optical depth of the second reionization
by a factor of $\sim 3$,
making the total Thomson optical depth much more significant.
Given the narrow allowed range of the first reionization epoch
between $z=15-16$ (Figure 8), 
our prediction of $\tau_e \sim 0.10\pm 0.03$
[where $\Delta tau_e=0.03$ is approximately estimated 
considering uncertainties on $C(z)$, $f_{\rm es}(z)$, $c^*(z)$ and
$\Omega_b$]
is fairly robust,
{\it unless the ionizing photon escape fraction from
the Pop III galaxies is extremely small} compared to
what is required at $z\sim 6$ to reionize the universe (Figure 7).
The implications for the CMB
(for an excellent recent review see Hu \& Dodelson 2002),
in particular, for polarization of the CMB
(e.g., Seljak 1997; Zaldarriaga 1997;
Kamionkowski \etal 1997) is significant.

We will follow the recent analysis by Kaplinghat \etal (2002)
on the detectability of polarization of the CMB.
Kaplinghat \etal (2002) show that the Microwave Anisotropy Probe
(MAP) satellite will be able to measure
$\tau_e$ to an accuracy of $0.02-0.03$ ($1\sigma$).
Thus it appears that
MAP will be able to distinguish the two reionization 
scenarios at about $2-3\sigma$ confidence level:
the standard reionization scenario 
where the universe is fully ionized at $z<6$
and fully neutral at $z>6$ (giving $\tau_e=0.027$)
versus the reionization scenario presented here
where the universe is fully ionized at $z<6$
and partially ionized at $z=6-20$ (giving $\tau_e=0.097$).
According to Kaplinghat \etal (2002)
Planck surveyor will probably be able to 
probe the detailed reionization history 
in addition to discriminating between the standard
reionization model and the current reionization model 
at a very high confidence level.

\subsection{Non-detection of Pop III Stars Locally}

There is a long history of search for 
Pop III stars 
(Bond 1970,1981;
Hills 1982;
Bessell \& Norris 1984,1987;  
Carr 1987;
Cayrel 1996;
McWilliam \etal 1995;
Ryan, Norris, \& Beers 1996;
Rossi, Beers, \& Sneden 1999;
Beers 2000;
Norris, Ryan, \& Beers 1997,2001;
Norris \etal 2002;
Hill \etal 2002;
Depagne \etal 2002)
or primeval galaxies
(Partridge 1974; Davis \& Wilkinson 1974).
If Pop III stars are as massive as suggested,
perhaps it is not surprising that   
no single Pop III star should have been found,
while over 100 metal-poor stars with $-4<[Fe/H]<-3$ have been
found (Cayrel 1996),
since all Pop III stars would end either as supernovae
or black holes.
The proposed model, unfortunately,
would mark the end of search for Pop III stars in the local
universe.

\subsection{Detectability of Galaxies Beyond $z=6$}

It is interesting to note, from Figure 9, that
there is a substantial redshift range
beyond the end redshift of the second reionization $z=6$,
where the IGM is already substantially ionized.
For example, $n_{\rm HI}/n_{\rm H}\sim 0.1-0.2$ 
at $z\sim 6.0-7.5$.
This feature of an extended redshift interval of 
high ionization appears to be generic.
This result would have interesting consequences
on the observability of galaxies beyond
the second reionization epoch.
Taking the result at face value,
for a galaxy at $z=6.5$,
the Str\"omgren sphere produced by the galaxy
will be a factor of about 2 larger than
that with the case where the galaxy is embedded in a completely neutral IGM.
This would help reconcile the seemingly 
conflicting observational claims:
on one hand, the universe appears to be 
reionized at $z\sim 6$ when
ionizing radiation background is seen 
to rise sharply  
(e.g., Fan \etal 2001; Becker \etal 2001, Barkana 2001; 
Cen \& McDonald 2002);
on the other hand,
\lya galaxies at redshift as high as $z\ge 6.5$ (Hu \etal 2002; 
Kodaira \etal 2003)
has been detected (see Haiman 2002 for an alternative
explanation).
The optical depth due to the damping wing
of neutral hydrogen outside the 
Str\"omgren is in general reduced by a factor of 
$(n_{\rm HI}/n_{\rm H})^{4/3}$
(Miralda-Escude \& Rees 1998;
Cen \& Haiman 2000;
Madau \& Rees 2000), 
which is a factor of $0.08$ for $n_{\rm HI}/n_{\rm H}=0.15$.
One prediction is that
the number of observable \lya galaxies
would quickly thin out beyond $z\sim 7$,
due to the combined effect of 
a rapid increase of neutral fraction beyond $z\sim 7$ (see Figure 9) and 
a decrease of the number of large \lya galaxies.
More detailed treatment of this important subject will be done
in a separate paper.
Observational campaign by several groups 
(Hu \etal 2002;
Rhoads \etal 2001;
Ajiki \etal 2002)
to detect high redshift \lya galaxies
may be able to probe the detailed
reionization structure near the end 
of the second cosmological reionization.

\subsection{Hydrogen 21cm Line due to Minihalos Prior to the Second Cosmological Reionization: A Test}

The possibility of probing the high-z universe
with hydrogen 21 cm line absorption or emission 
has been suggested and investigated in 
various contexts by many authors (Hogan \& Rees 1978;
Scott \& Rees 1990;
Subramanian \& Padmanabhan 1993;
Kumar, Padmanabhan, \& Subramanian 1995;
Bagla, Nath, \& Padmanabhan 1997;
Madau, Meikien, \& Rees 1997;
Shaver \etal 1999;
Tozzi \etal 2000;
Carilli, Gnedin, \& Owen 2002;
Iliev \etal 2002;
Furlanetto \& Loeb 2002).
The mean, neutral medium expanding with the Hubble flow
at redshift $z$ would produce an optical depth 
for the hydrogen 21 cm resonant absorption 
is (Shklovsky 1960)
\begin{equation}
\bar\tau(z) = {n(z) \over H(z)} {g_2\over g_1(g_2+g_1)} A_{21} {c^3 \over 8\pi\nu^3} {h\nu\over kT_{\rm sp}} = 8.5\times 10^{-4} ({T_{\rm sp}\over 200{\rm K}})^{-1} ({\Omega_{\rm b} h^2\over 0.02})({\Omega_{\rm M} h^2\over 0.15})^{-1/2}({1+z\over 8})^{3/2}
\label{taueq}
\end{equation}
where $H(z)$ is the Hubble constant at redshift $z$;
$n(z)$ is the mean atomic hydrogen number density at $z$;
$g_1=1$ and $g_2=3$  are the statistical weights of the lower and
upper hyperfine levels;
$\nu$ is the frequency of the hydrogen 21cm line;
$T_{\rm sp}$ is the spin temperature of the atomic hydrogen;
$\nu$ is the frequency of the 21 cm line;
$c$ is the speed of light;
$k$ is the Boltzmann's constant;
$A_{21}$ is the spontaneous decay rate of the 
hyperfine transition of atomic hydrogen;
$\Omega_M$ and $\Omega_b$ are total matter and baryonic matter density
at $z=0$ in units of closure density;
$h=H_{\rm 0}/100$km/s/Mpc, where $H_{\rm 0}$ is the Hubble constant
at $z=0$.
Clearly, the uniform medium would only cause a very modest
absorption at $z\sim 6-15$.

Minihalos would cast significantly larger optical depth.
For 21cm absorption by atomic hydrogen in virialized minihalos 
we may rewrite Equation (30) approximately as:
\begin{equation}
\tau_h(z) = {178 n(z) r_v \over \sigma_v} {g_2\over g_1(g_2+g_1)} A_{21} {c^3 \over 8\pi\nu^3} {h\nu\over kT_{\rm sp}}
\label{taueq}
\end{equation}
where $r_v$ and $\sigma_v$ are the virial radius and
velocity dispersion within the virial radius, respectively;
the factor $178$ is the relative overdensity of matter
within the virial radius.
Using relations $\sigma_v=GM_v/2r_v$, 
$M_v=178 (4\pi/3) \rho(z) r_v^3$ [where
$\rho(z)$ is the mean total density at $z$]
and the definition of Hubble constant,
we can transform Equation (31) to

\begin{equation}
\tau_h(z) = 2\sqrt{178} \bar\tau(z) = 0.023 ({T_{\rm sp}\over 200{\rm K}})^{-1} ({\Omega_b h^2\over 0.02})({\Omega_M h^2\over 0.15})^{-1/2}({1+z\over 8})^{3/2}.
\label{taueq}
\end{equation}
\noindent
We see that gaseous minihalos with $T_{\rm sp} \sim 10^2-10^3$,
{\it if they exist},
could produce
significant optical depths at 21 cm 
at a level of $0.01-0.1$ (Furlanetto \& Loeb 2002; see also Carilli,
Gnedin \& Owen 2002 for a numerical treatment). 
Note that the spin temperature $T_{\rm sp}$ would
be bracketed by the cosmic microwave background temperature
and the kinetic temperature (approximately the virial temperature)
of a minihalo.

Furlanetto \& Loeb (2002) have pointed out 
that the 21 cm forest produced by minihalos 
prior to cosmological reionization
will be detectable by the next generation of low-frequency
radio telescopes, such as 
Low Frequency Array (LOFAR) and 
Square Kilometer Array (SKA).
However, this will only be possible, {\it if the minihalos
could retain their gas}, as they would 
in the conventional reionization scenario.
In the reionization scenario presented here,
as indicated in Figure 14,
no significant amount of gas 
will be able to accumulate in minihalos 
throughout the second reionization period.
Consequently, in our scenario, 21 cm forest lines,
which would otherwise be produced by minihalos without the first reionization,
have been largely wiped out.
It is noted that 21 cm absorption 
measurements would require bright background radio sources
to be present at very high redshift.

Similarly, all hydrogen 21 cm emission line forest due to 
minihalos, as many as $\sim 100$ lines per unit redshift 
at $z\sim 9$, as predicted by Iliev \etal (2002) in the standard
reionization scenario,
will be absent in the present reionization scenario. 
In addition, the 21 cm emission spatial
and spectral signature at $z\ge 6$ 
(Madau \etal 1997; Tozzi \etal 2000)
will also be significantly altered in the sense
that signals due to minihalos will be removed.

Hydrogen 21 cm observations, either in absorption or emission,
will be a definitive test to distinguish between the
two reionization scenarios.

\subsection{Metal Absorption Lines at $z>6$: A Probe of the Second Reionization Process and Initial Metal Enrichment History}

Oh (2002) pointed out that some metal absorption lines 
may be observable at high redshift.
He identified OI ($1302\AA$) and SiII ($1260\AA$)
lines, longward of the hydrogen \lya line,
as the best candidates
for tracing neutral hydrogen due to the proximity of their
ionizational potentials to the hydrogen ionization potential.
Since both these metal species trace closely
the neutral hydrogen density, OI and SiII absorption forest
may provide a way to probe the reionization history.
Two possible utilizations are noted here.
First, both OI and SiII absorption forests
are expected to thin out at $z\le 7$,
which could provide a direct probe of 
the end of the second reionization process.
We note that, for IGM which is primarily metal enriched
by Pop III stars,
the overabundances of $\alpha$-elements such as oxygen 
and silicon relative to the solar value (Heger \& Woosley 2002)
would greatly increase 
the optical depth of the oxygen OI and SiII absorption forests,
at the apparent low metallicity (defined by $[Fe/H]\sim 10^{-3}$).
This would substantially increase the number of observable
OI and SiII absorption lines predicted by Oh (2002).
Second, the relative optical depths of OI and SiII absorption lines
may shed light on the relative abundance of O to Si,
giving useful information on synthesis in Pop III stars
(Qian \& Wassurberg 2000,2002).
In addition, gradually increasing contribution from Pop II 
supernovae (type II) to the IGM metals with time may also show up
in the redshift evolution 
of the relative optical depths of OI and SiII absorption lines,
possibly providing information on the overall metal enrichment
history of the universe at those early times.

\subsection{Comparisons with Some Other Works}

The author was made aware of an earlier paper by 
Wyithe \& Loeb (2002), who investigated the reionization
histories due to stars and quasars.
They assume that minihalos do not form stars at
the redshift of interest but
systematically explore the 2-d parameter space spanned
by the ionizing photon escape fraction and the transition
redshift from a VMS IMF and a normal IMF.
They pointed out that, for some restricted region
in the parameter space, hydrogen and/or helium II 
may be reionized twice.
The primary differences between their study
and the current study are three-fold.
First, we show that \h2 cooling thus star formation in
minihalos takes place unimpeded, as long as gas is able to 
accrete onto them,
whereas they do not consider star formation in minihalos.
Second, we constrain the ionizing photon escape
fraction directly using \lya forest observations.
As a consequence, we show that it seems 
inevitable that the universe will be reionized
first by Pop III stars either in minihalos or in large halos
and second by Pop II stars.
Finally, we use a new computational method to follow self-consistently
the evolution of the entire IGM, including HII regions,
HI regions and partially ionized regions.
The two studies are, however, complementary 
in their different approaches.

Venkatesan, Tumlinson, \& Shull (2003)
also examined the consequences of Pop III metal-free stars 
on the reionization of the universe.
They indicated the possibility of reionizing He II twice
with an incomplete first attempt of reionizing He II.

Mackay, Bromm, \& Hernquist (2002) 
presented a three-phase reionization picture.
With a top-heavy VMS IMF, they pointed out 
that the first transition from molecular cooling 
phase to atomic cooling phase
occurs at $z\sim 30$, when their
demand of 10 Lyman-Werner photons per baryon is met.
The proposed first transition in Mackay \etal (2002)
is dictated by the lack of \h2 cooling in minihalos.
We pointed out here positive feedback and as a result 
there is no such transition.
Mackay \etal (2002) find that 
at $z=15-20$ the IGM has been enriched to a level
of $10^{-3.5}\zsun$ that the second transition
to a normal IMF occurs.
Their assessment of the transition epoch from Pop III to Pop II
is consistent with but at a slightly
higher redshift than our calculation.
Their universe, however, is not reionized until $z\sim 6$.
It is noted in passing that Mackay \etal (2002)
pointed out a novel Pop II.5 stellar population
associated with cooling shells of Pop III supernovae.
In summary, the primary differences between this study and that by 
of Mackay \etal (2002) 
are two-fold.
First, we indicated that the positive feedback mechanisms on \h2 formation
allow star formation to occur in minihalos
throughout the first reionization period up to $z\sim 13.2$.
Second, we constrain the ionizing photon escape fraction
by the direct \lya forest observations to eliminate this major uncertainty.

\subsection{Uncertainties}

Perhaps the most uncertain in the chain of derivation 
is the IMF of the Pop III stars.
However, most of the main conclusions, including
positive feedback of Pop III star formation
and two times of cosmological reionization,
are likely to hold for any sufficiently top-heavy IMF.
While the simulations with sub-solar mass
resolution (Abel \etal 2002; Bromm \etal 2002)
have clearly shown the formation of Pop III VMS,
it is yet unclear what the angular momentum 
transport mechanisms are.
It is possible, in principle, that hydrodynamic
processes, included in the quoted simulations,
could transport the unwanted angular momentum outward,
as the simulators have advocated.
If local star formation is a guide,
one could imagine that very massive Pop III stars
may form in binaries or multiples, in which case
angular momentum removal would be easily achieved.
It is also intriguing to note that,
if we extrapolate the observed
ratio of super massive black hole to bulge 
mass (e.g., Tremaine \etal 2002),
then one would expect
to find a black hole of mass $\sim 100-1000\msun$
at the center of each minihalo of mass
$10^5-10^6\msun$.
Thus, perhaps one should not be too surprised
that nature has managed to 
form a compact object (star) of mass $\sim 100\msun$
formed at the center of a minihalo, 
if the quoted observations are any empirical guide;
it may be argued,
angular momentum-wise, that it is a less stringent task to 
collapse a higher density gas cloud at higher redshift
to form a star than a lower density gas cloud at lower
redshift to form a black hole, as observed.

We have ignored the possible contribution from quasars to
the second reionization process.
If the emission rate of ionizing photons from quasars
is proportional to the star formation rate,
as indicated by low redshift observations 
(Boyle \& Terlevich 1998; Cavaliere \& Vittorini 1998),
then, to zero-th order, all that is needed is to 
renormalize $f_{\rm es}$ such that
the second reionization completes at $z\sim 6$
and results would remain largely unchanged.

\section{Conclusions}

The conclusions of this paper consist of two sets.

I) We put forth two new mechanisms for generating
a high X-ray background during the Pop III era.
namely, X-ray emission from the cooling energy of 
Pop III supernova blast waves and 
from miniquasars powered by Pop III black holes.
We show, consequently, \h2 formation in the cores of minihalos
is significantly induced,
more than enough to compensate for destruction by Lyman-Werner photons
produced by the same Pop III stars.
In addition, another, perhaps dominant process for 
producing a large number of \h2 molecules in relic HII
regions created Pop III galaxies, first pointed out RGS,
is quantified here.
We show that \h2 molecules produced 
by this process may overwhelm the Lyman-Werner photons 
produced by the stars in the same Pop III galaxies.
As a result, the Lyman-Werner background may never be able to
build up to affect \h2 molecules in minihalos.
In combination, we suggest that 
cooling and hence 
star formation in minihalos can continue to take place
largely unimpeded throughout the first reionization period,
as long as gas is able to accumulate in them.

II) We show that the intergalactic medium is likely to have been
reionized twice, first at $z=15-16$ and second at $z=6$.
Under a very conservative assumption for the star formation
efficiency in minihalos with \h2 cooling,
the first reionization may be attributable largely to
Pop III stars in large halos;
in this case, processes promoting \h2 formation in the
cores of minihalos such as mentioned above in (I)
do not really matter.
In contrast, if star formation
efficiency in minihalos is not more than
a factor of ten less efficient than in large halos,
then Pop III stars in minihalos may be largely responsible
for the first reionization.
In either case, it seems likely that
Pop III stars in minihalos make a large
contribution to the first phase of the metal enrichment
of the intergalactic medium approximately up to 
a level where the transition from Pop III stars to Pop II
stars occurs.

This apparent inevitability of twice reionizations
is reached, based on a joint constraint by
two observational facts:
1) the universe is required to be reionized at $z\sim 6$
and 2) the density fluctuation in the universe
at $z\sim 6$ is well determined by 
the same small-scale power traced by the 
\lya forest observed and well measured at $z\sim 3$.
As a result, the product of star formation efficiency
and ionizing photon escape fraction from galaxies
at high redshift is well constrained, 
dictating the fate of the cosmological reionization process.

The prolonged reionization and reheating history of the 
IGM is more complicated than usually thought.
Because both cooling and recombination time scales
are much shorter than the Hubble time at the redshift
in question,
we devise a new, improved computational method to
follow the evolution of the IGM in all phases,
including HII regions, HI regions and partially
ionized regions.
The overall reionization process may be 
separated into four joint stages.

(1) From $z\sim 30$ to $z\sim 15-16$,
Pop III stars gradually heat up and ionize
the IGM and the first reionization occurs at the end
of this stage, when the mean temperature of the IGM reaches
$\sim 10^4$K.

(2) The first stage is followed by a brief period
with a redshift interval of order $\Delta z\sim 1$,
where the IGM stays completely ionized due to sustained ionizing
photon emission by forming Pop III stars.
During this period, the temperature of the IGM is maintained
at $\sim 10^4$K.

(3) The transition from Pop III stars to Pop II stars
sets in at the beginning of this stage ($z\sim 13$).
The abruptly reduced (by a factor of $\sim 10$) 
ionizing photon emission rate
causes hydrogen to rapidly recombine and 
the universe once again becomes 
opaque to \lya and Lyman continuum photons,
marking the second cosmological recombination.
From this time until $z=6$
Compton cooling by the cosmic microwave background
and photoheating by the stars self-regulate 
the Jeans mass and the star formation rate.
The mean temperature of the IGM
is maintained at $\sim 10^4~$K from $z\sim 13$ to $z=6$.
Meanwhile, recombination and photoionization 
balance one another such that
the IGM stays largely ionized
during this stage with $n_{\rm HII}/n_{\rm H}\ge 0.6$.
Most of the star formation in this period occurs in
large halos with dominant atomic line cooling.

(4) At $z=6$, the global star formation rate 
again surpasses the global recombination 
rate, resulting in the second reionization of the universe.
The second reionization is predominantly 
due to stars formed in halos where atomic line cooling is efficient.


There is a wide range of interesting implications 
from this new reionization picture presented here.
To highlight a few: 

$\bullet$ The Pop III stars
enrich the IGM with a metallicity of $\sim 10^{-3}\zsun$ at $z\sim 13-15$.

$\bullet$ The magnetic field originating from massive stars
could pollute the IGM with a large-scale coherent
field ($l\sim 100$kpc comoving) of order $\sim 10^{-9}~$G at $z\sim 13-15$.

$\bullet$ The number density of Pop III massive black holes
($10-300\msun$) is comparable to that of globular clusters
and should seed later structure formation.


$\bullet$ Direct detection of Pop III hypernovae/supernovae 
are very difficult but a systematic search for 
gravitational lensing magnified Pop III hypernovae/supernovae 
targeted at massive clusters of galaxies
may turn out to be fruitful; SIRTF may be able 
to detect them. 
SIRTF may also be able to detect some of the large Pop III galaxies.
JWST should be able 
to detect both large Pop III galaxies and Pop III hypernovae.

$\bullet$ The Thomson scattering optical depth is increased to
$0.10\pm 0.03$ (compared to $0.027$ 
for the case of only one rapid reionization at $z=6$),
which will have significant implications on the
polarization observations of the cosmic microwave background.
Upcoming MAP results should be able to distinguish 
between these two scenarios.

$\bullet$ Under the present scenario it would not have been
a surprise that no Pop III stars have been found in the local universe.

$\bullet$ The IGM, while opaque to \lya photons,
is highly ionized ($n_{\rm HI}/n_{\rm H}\sim 0.1-0.2$)
up to redshift $z\sim 7$.
This may reconcile the observation indicating 
the second reionization
at $z\sim 6$ with the detection of \lya galaxies
at $z\ge 6.5$.

$\bullet$ Finally, in contrast to the conventional reionization scenario,
21 cm lines prior to $z=6$,
which would otherwise be produced by minihalos 
in the absence of the first cosmological reionization,
are predicted not to exist.
This would provide a definitive test of the scenario.

\acknowledgments
I thank Nick Gnedin, Zoltan Haiman and Martin Rees for
stimulating discussion,
Rennan Barkana, Ave Loeb and Jerry Ostriker for helpful comments
and Rupert Croft for information on  
constraints on cosmological models
placed by \lya forest observations.
This research is supported in part by NSF grant AST-0206299.

\begin{appendix}

\section{Magnification of Pop III Sources by Gravitational Lensing}

We determine the fraction of area
in the background source plane at redshift $z_S$, 
that is magnified by a factor of $\ge \mu$
by foreground objects,
primarily clusters of galaxies.
The magnification regime in which we are interested is $\mu\ge 10$.
We will first derive a general
relationship between the magnification ($\mu$)
and source plane coordinate ($x$): $\mu \propto 1/x$ for
any singular density profile.
Then, we will use the singular isothermal sphere thanks to
its ease of analytic treatment,
which has the same $\mu\propto 1/x$ relation, 
to model clusters of galaxies 
to compute the gravitational lensing cross section at
the high $\mu$ end.
The computed lensing cross section is cross-checked by
observations to ensure consistency.

\begin{figure}
\plotone{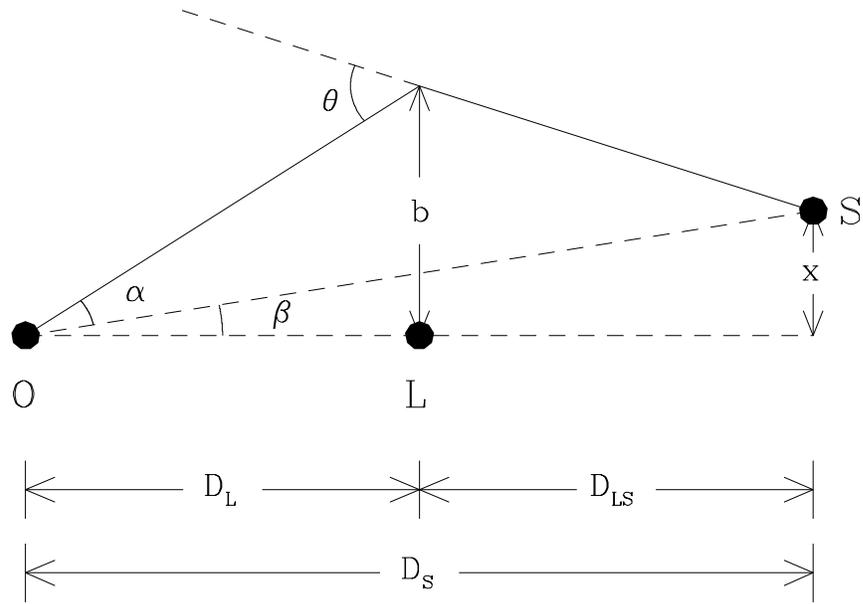}
\caption{
shows the gravitational lensing optics.
$O$, $L$ and $S$ are the observer, the lens and the source, respectively.
$D_L$, $D_S$ and $D_{LS}$ are angular diameter distance
from observer to the lens, from observer to the source,
and from the lens to the source, respectively.
$b$ and $x$ are the impact distance and source plane coordinate,
respectively.
$\theta$ is the light bending angle when passing by the lens at
impact distance $b$.
}
\label{lens}
\end{figure}

We express the magnification $\mu$
as a function of $x$, the source plane proper coordinate (Fig 13).
It is well known that 
$\mu \approx 1/x$ in the small $x$, high $\mu$ limit,
for both a point mass lens (Vietri \& Ostriker 1983)
and a singular isothermal sphere lens
(SIS; Turner, Ostriker \& Gott 1984).
For simplicity we will consider only {axisymmetric lenses}. 
We will
generalize this property of $\mu$ in the small $x$, high $\mu$ limit
to any lens system as long as
the lensing bending angle, $\theta$,
as a function of proper impact parameter, $b$, does not increase
as fast as $b$.
It is noted that 
$\theta\propto b$ for a uniform mass sheet,
$\theta=\;$constant for a SIS,
and $\theta\propto 1/b$ for a point mass.
From Figure 15 we can readily write down the lens equation as:
\begin{equation}
\alpha + \beta = {b\over D_L},~~~ \beta={x\over D_S},~~~\alpha D_S=\theta D_{LS}.
\end{equation}
\noindent 
The above equation relates
the angular position of the image, $\alpha+\beta$, 
to the position of the source, $\beta$ 
(both with respect to some reference direction)
through the bending angle, $\theta$.
$D_S$, $D_L$ and $D_{LS}$ are the angular diameter distance 
between the observer and the source,
between the observer and the lens,
and between the lens and the source, respectively.
$b$ is the proper impact parameter relative to the lens center.
Expressing the bending angle $\theta$ as $\theta \equiv \theta_0 (b/b_0)^m$ ($m<1$),
we can rewrite the above lens equation as:
\begin{equation}
\theta_0 (b/b_0)^m = b D_S/(D_L D_{LS}) - x/D_{LS}.
\end{equation}
\noindent 
In the limit $x\rightarrow 0$, the solution for the impact parameter
is $b=b_0[b_0D_S/(\theta_0 D_L D_{LS})]^{1/(m-1)}$ (Einstein radius),
and $dx/db = (1-m) D_S/D_L$.
It is noted that the requirement $m<1$ is to guarantee
a solution, i.e., the existence of 
an Einstein radius. 
Then, it is straightforward to show that,
at small $x$ limit, the magnification is
\begin{equation}
\mu(x)={b\over x}{db\over dx} ({D_S\over D_L})^2={K\over x}
\end{equation}
\noindent 
with the constant $K$, independent of $x$, being
\begin{equation}
K=({b_0 D_S\over\theta_0D_LD_{LS}})^{1/(m-1)}{D_S\over D_L}{b_0\over 1-m}.
\end{equation}
\noindent 
We note that, for any (centrally) 
singular density profile,
$m$ is less than unity.
It has been shown 
by Syer \& White (1996) 
and Subramanian, Cen, \& Ostriker (2000)
that
the density profile in the inner regions of dark matter halos,
which are formed through hierarchical gravitational clustering/merging
in the conventional Gaussian structure formation models,
is $\rho (r) \propto r^{-3(3+n)/(5+n)}$,
where $n$ is the linear power spectrum index at the relevant scales.
For the scales of interest, we have $n\sim -2$ to $-1$
for cold dark matter models, giving
$\rho (r) \propto r^{-1.5}$ to  $\rho (r) \propto r^{-1.0}$.
These inner slopes of halos were
borne out in N-body simulations
(Navarro, Frenk, \& White 1997;
Moore \etal 1999).
Thus, we should expect $\mu\propto 1/x$ at small $x$
for halos formed in hierarchical structure formation models,
including those of galaxy cluster size.
Moreover, cooling and subsequent condensation of baryons in the 
centers of halos may further steepen the density profiles
in the inner regions.

Having deduced the universal $\mu\propto 1/x$ relation at the small $x$ end,
we will now carry out
further calculations by adopting the SIS
model,
because of its analytical simplicity 
and because of its astrophysical relevance 
as shown in many statistical studies (Gott \& Gunn 1974;
Tyson 1983;
Turner \etal 1984;
Hinshaw \& Krauss 1987;
Narayan \& White 1988;
Wu 1989;
Fukugita \& Turner 1991;
Mao 1991).
This is further
justified if $\mu$ is normalized at a somewhat lower value,
where some data from current observations exist.
We note that, for individual multiply imaged observed quasars,
it is clear that one needs to include ellipticities
of the lenses for realistic modeling.
For example, one cannot produce quadruples with an
axisymmetric lens (Narayan \& Grossman 1989;
Blandford \etal 1989).
However, for our purpose of calculating the
magnification cross section at moderate-to-high range $\mu \sim 10-100$,
the assumption of axisymmetry of lenses should be adequate.
For a SIS the bending angle, $\theta_0$, 
conveniently independent of impact parameter,
is (Turner \etal 1984)
\begin{equation}
\theta_0= {4\pi}({\sigma_{||}\over c})^2,
\end{equation}
\noindent 
where $c$ is the speed of light and $\sigma_{||}$
is the line of sight velocity dispersion of the SIS lens.
With equation (A5) we can solve equation (A2) with the following solution
at the $x\rightarrow 0$ limit:
$b={\theta_0D_L D_{LS}\over D_S}$ and 
${db\over dx}={D_L\over D_S}$.
Inserting this solution into equation (A3) yields
the small $x$ limit of the $x-\mu$ relation for a SIS lens:
\begin{equation}
x(\mu)={4\pi\over \mu}({\sigma_{||}\over c})^2 {D_{LS}}.
\end{equation}
\noindent 
A SIS would subtend a solid angle in the source plane $\Delta\Omega$ 
within which the luminosity of a source is magnified by $\ge\mu$:
$\Delta \Omega= {\pi x^2(\mu)\over D_S^2}$.
Thus, the total probability of a {\it random} 
source at $z_S$ of
being magnified by $\ge\mu$ 
can be obtained by adding up all the foreground lenses.
This is done by integrating $\Delta\Omega /4\pi$
over lenses of all masses and over all redshifts up to the source 
redshift, $z_S$, resulting in a double integral:
\begin{equation}
P_{random}=\int_0^\infty \int_0^{z_S} {\pi x^2\over 4\pi D_S^2}
    n(M_A,z) 4\pi r^2 {dr\over dz} dz dM ,
\end{equation}
\noindent 
where $n(M_A,z)$ is the mass function of halos at redshift $z$.
To make the subsequent calculation more analytically tractable,
we assume 
\begin{equation}
n(M_A,z)= (1+z)^{-w} g(M_A),
\end{equation}
\noindent 
where
\begin{equation}
g(M_A)=(4\times 10^{-5}/M_*) (M_A/M_*)^{-1}\left[1+(M_A/M_*)^{-1}\right] \exp(-M_A/M_*)~h^{3}\hbox{Mpc}^{-3}
\end{equation}
\noindent 
is the differential cluster mass function at $z=0$, taken
from Bahcall \& Cen (1993). 
$M_A$ is the observed cluster mass within the Abell radius
($r_A=1.5h^{-1}$Mpc) and $M_*=1.8\times 10^{14}~h^{-1}$M$_\odot$.
The term $(1+z)^{-w}$ is intended to approximately describe
the evolution of the overall cluster mass function with redshift.
The cluster mass, $M_A$,
defined within the Abell radius, $r_A$,
may be related to $\sigma_{||}$ for the SIS model by
\begin{equation}
\sigma_{||}^2={G M_A\over 2 r_A},
\end{equation}
\noindent 
where $G$ is the gravitational constant.
Combining equations (A6) and (A10) indicates that
$x^2 \propto M_A^2$.
It is instructive 
to examine the 
behavior of the
following term, $x^2 g(M_A)$, which 
contains all the dependence of 
the integrand of the integral in equation (A7) on $M_A$.
Since $x^2g(M_A)$ is constant at the low $M_A$ end
and $x^2g(M_A)\propto (M_A/M_*) \exp(-M_A/M_*)$ at the high $M_A$ end,
the integral over $M_A$ in equation (A7) is 
convergent on both ends and 
dominated by massive clusters near
the exponential downturn at about $M\sim 1\times 10^{15}M_\odot$.
Consequently, for our purpose, we need only to be accurate on
the high mass end near the exponential downturn for $g(M_A)$ 
and its redshift distribution characterized by parameter $w$.
Substituting equations (A6, A8, A9, A10) into equation (A7)
[and making use of the simple relations in an $\Omega_0=1$ universe:
$D_L=R_H(1-1/\sqrt{1+z_L})$,
$D_S=R_H(1-1/\sqrt{1+z_S})$,
$D_{LS}=R_H(1/\sqrt{1+z_L}-1/\sqrt{1+z_S})$
and ${dr(z)\over dz} = {1\over 2} R_H (1+z)^{-3/2}$]
yields
\begin{eqnarray}
P_{random}&=&{2B\pi^3 G^2 R_H^3\over c^4 r_A^2 \mu^2}\left(1-{1\over\sqrt{1+z_S}}\right)^{-2} \int_0^\infty M_A^2 g(M_A) dM \nonumber\\ 
&&\times \int_0^{z_S} \left(1-{1\over\sqrt{1+z}}\right)^{2} \left({1\over\sqrt{1+z}}-{1\over\sqrt{1+z_S}}\right)^{2} (1+z)^{-3/2-w} dz \nonumber\\
&=&{1.0\times 10^{-5} B \pi^3 G^2 R_H^3 M_*^2\over c^4 r_A^2 \mu^2}\left(1-{1\over\sqrt{1+z_S}}\right)^{-2} \int_0^\infty t (1+t^{-1})\exp(-t) dt  \nonumber\\
&&\times \int_0^{z_S} \left(1-{1\over\sqrt{1+z}}\right)^{2} \left({1\over\sqrt{1+z}}-{1\over\sqrt{1+z_S}}\right)^{2} (1+z)^{-3/2-w} dz.
\end{eqnarray}
\noindent 
All constants in equation (A11) are in c.g.s. units,
except $R_H$, which is in Mpc.
Note that we have also inserted a constant $B$
in equation (A11), which serves to absorb the uncertainties
due to other factors 
which either cannot be accurately treated here
or are unknown, including background cosmology, 
deviations of density profiles from singular isothermal spheres,
gravitational lensing due to
other astronomical objects
and uncertainties in the observed cluster mass function.
The normalization of $B$ will be set by comparing to
observations at $\mu>2$.
Now expanding all the constants
and integrating equation (A11) with respect to $t$ ($\equiv M_A/M_*$),
we obtain
\begin{equation}
P_{random}={0.036B\over \mu^2} (1-{1\over\sqrt{1+z_S}})^{-2} I(z_S,w),
\end{equation}
\noindent where 
$I(z_S,w) \equiv \int_0^{z_S}
(1-{1\over\sqrt{1+z}})^2 
({1\over\sqrt{1+z}}-{1\over\sqrt{1+z_S}})^2 (1+z)^{-3/2-w} dz$.
The integral $I(z_S,w)$ can be done analytically 
(but the resultant expression is quite lengthy)
and here we just give the final numbers for specific $z_S$ and $w$.
For $z_S=17.0$,
$P_{random}=(1.1\times 10^{-3}, 6.8\times 10^{-4}, 
4.2\times 10^{-4}, 2.1\times 10^{-4})B/\mu^2$,
for $w=(0.0,0.5,1.0,2.0)$, respectively. 
It shows that $P_{random}$ only weakly depends on $w$,
because the integral is dominated by the moderate redshift
range $z\sim 0.5-1.0$.

It is justified to only consider the large splitting events since
the lensing cross section is dominated by massive clusters which
give rise to large splittings ($>1^{''}$).
For the sake of concreteness we will adopt $w=0.5$, which
is consistent with the relatively wild evolution of cluster density
up to redshift about unity (e.g., Bahcall, Fan, \& Cen 1997). 
As we will show below, the final results ($P_{clust}$)
turn out to be extremely weakly dependent on $w$.
Furthermore, we take $B=3.4$ to normalize $P_{random}$:
\begin{equation}
P_{random}(\mu)={2.0\times 10^{-3}\over \mu^2},
\end{equation}
\noindent 
which gives $P_{random}(\mu=2) = {4.0\times 10^{-4}}$.
Since we are integrating the observed clusters 
to obtain the lensing probability and the same
clusters are also responsible for observed gravitational
lensing events, we can make a consistency check.
For the Hewitt \& Burbidge (1989) quasar catalog of
4250 quasars there are two confirmed multiple image lens
systems with splitting $>1^{''}$, which corresponds to 
multiple lensing probability of $4.7\times 10^{-4}$($=2/4250$),
i.e., $P=4.7\times 10^{-4}$ for $\mu>2$ 
(in the singular isothermal sphere case).
Thus, our adopted normalization is consistent with 
the observed value of $P=4.7\times 10^{-4}$ for $\mu>2$.
Considering
possible selection biases (Turner 1980; Cen \etal 1994a; Kochanek 1995) 
and the countervailing effect that the observed quasar sample
is at significantly lower redshift than $z=17$ considered here,
we argue that the adopted normalization is reasonable.
From this we see that a {\it random} source at $z_S=17$
has a rather small probability of $\le 10^{-3}$ being very strongly lensed.

We will now examine the case where we do not 
sample the sky randomly, rather observe 
a set of selected regions centered on massive clusters.
The cross section 
for magnification $\ge\mu$ in the source plane
is $\pi x^2(\mu)$, 
so the solid angle within which 
the sources will be magnified by $\ge\mu$
is $\pi x^2(\mu)/D_s^2$.
Integrating $\pi x^2(\mu)/D_s^2$
over a pre-selected set of clusters with mass $M_A\ge M_{lim}$ 
in the redshift range
$z=z_1$ to $z_2$ 
and dividing the integral by 
the number of clusters pre-selected
gives the mean solid angle 
for source magnification $\ge\mu$.
\begin{equation}
\langle \Omega(>\mu)\rangle ={\int_{M_{lim}}^\infty B^\prime g(M_A) dM\int_{z_1}^{z_2} {\pi x^2\over D_S^2} (1+z)^{-w} 4\pi r^2 {dr\over dz} dz \over
N_{cl}(z_1,z_2,M_{lim})},
\end{equation}
\noindent
where $N_{cl}(z_1,z_2,M_{lim})\equiv \int_{M_{lim}}^\infty g(M_A) dM\int_{z_1}^{z_2} (1+z)^{-w} 4\pi r^2 {dr\over dz} dz$ 
is the total number of clusters selected
(a normalization factor).
Dividing the mean solid angle 
for source magnification $\ge\mu$
by the field of view
yields the mean probability of {\it a source} in such selected fields 
which will be magnified by $\ge\mu$:
\begin{equation}
P_{clust}(\mu)={\langle \Omega (>\mu)\rangle \over \Omega_{FOV}},
\end{equation}
\noindent
where $\Omega_{FOV}$ is a telescope's field of view. 
As an example, let us take $M_{lim}=1\times 10^{15}M_\odot$,
$z_1=0.0$ and $z_2=0.4$, which yields
$N_{cl}(0.0,0.4,1\times 10^{15})=(3020,2669,2362,1852)$,
for $w=(0.0,0.5,1.0,2.0)$, respectively,
in an $\Omega_0=1$, $H=50$km/s/Mpc universe.
Similar to equation (A11),
we can integrate equation (A14) 
for $z_S=6.0$, $z_1=0.0$, $z_2=0.4$
and $M_{lim}=1.0\times 10^{15}M_\odot$ and
the result is:
$P_{clust}(\mu)=(6.7\times 10^{-2}, 6.7\times 10^{-2}, 6.8\times 10^{-2}, 6.9\times 10^{-2}) {1\over \mu^2} ({\Omega_{FOV}\over 25~\hbox{arcmin}^2})^{-1}$.
Although an adjustment parameter
$B^\prime$ in equation (A14), which is different from
$B$ in equation (A11) to reflect the fact that
the uncertainties involved are somewhat different
(for example, since the probability is normalized per cluster,
the uncertainty in the amplitude of the observed  cluster mass function
cancels out),
for clarity and simplicity we adopt $B^\prime=B=3.3$.
The result indicates that $P_{clust}(\mu)$ is nearly independent of
$w$, unlike $P_{random}$, which depends on $w$, giving
\begin{equation}
P_{clust}(\mu)={6.7\times 10^{-2}\over \mu^2}\left({\Omega_{FOV}\over 25~\hbox{arcmin}^2}\right)^{-1}.
\end{equation}
\noindent 
Comparing $P_{clust}$ with $P_{random}$, 
for the case with $w=0.5$, for example, we see that one gains a
factor of $34$ in the cluster centered survey
compared to a random survey,
for a field of view of $\Omega_{FOV}=25$~arcmin$^2$.
The relative gain decreases with increasing size of the field of view,
and the two probabilities become equal at $\Omega_{FOV}=$838~arcmin$^2$,
above which the calculation of $P_{clust}$ is no longer valid
due to multiple clusters in a single field.
Note that,
although the mean probability of a source within 
the field of view is inversely proportional to $\Omega_{FOV}$,
the total number of magnified sources per field of view
is $P_{clust} \Omega_{FOV} \Sigma_{src}$ (where $\Sigma_{src}$
is the surface density of source $src$, which is either $gal$ or
$SN$ in this study), which is independent of $\Omega_{FOV}$.
The physical reason for having a relative increase in 
the probability of magnified sources 
is that the clusters of galaxies are rare targets
and hence a {\it random} field of view has a small
probability of intersecting a cluster.
We note that $P_{random}$ is still valid even 
in the case of multiple clusters in a single field.
However, the calculation of $P_{random}$ breaks down
when multiple clusters are {\it precisely} aligned along
the line of sight.
But such cases are likely to be negligibly few.
We further note that the spatial clustering of galaxy clusters 
would further boost the gain of $P_{clust}$ 
over $P_{random}$, but we will not consider
this effect here.

\end{appendix}

\end{document}